\documentclass[journal,twoside,web]{ieeecolor}
\usepackage{generic}
\usepackage{cite}
\usepackage{amsmath,amssymb,amsfonts}
\usepackage{algorithm,algorithmic}
\usepackage{graphicx}
\usepackage{textcomp}

\usepackage{gensymb}
\usepackage{hyperref}
\usepackage{todonotes}
\usepackage{multirow}
\usepackage{booktabs}

\usepackage{tikz}
\usepackage{pgfplots}
\usepackage{pgfplotstable,booktabs}
\usetikzlibrary{pgfplots.groupplots}
\usepgfplotslibrary{dateplot,fillbetween}
\usepackage{pgf}
\usepgfplotslibrary{colorbrewer}
\def\BibTeX{{\rm B\kern-.05em{\sc i\kern-.025em b}\kern-.08em
    T\kern-.1667em\lower.7ex\hbox{E}\kern-.125emX}}
\usepackage{titlesec}
\titlespacing*{\section}
{0pt}{2.3ex}{.6ex}
\titlespacing*{\subsection}
{0pt}{2.1ex}{.6ex}
\newtheorem{assumption}{Assumption}

\newtheorem{lemma}{Lemma}
\newtheorem{corollary}{Corollary}
\newtheorem{definition}{Definition}
\newtheorem{remark}{Remark}

\DeclareMathOperator{\Han}{\mathfrak{H}}
\DeclareMathOperator{\R}{\mathbb{R}}
\DeclareMathOperator{\E}{\mathbb{E}}
\DeclareMathOperator{\tr}{\text{tr}}
\DeclareMathOperator*{\argmin}{arg\,min}

\newcommand{\change}[1]{\textcolor{black}{#1}}
\newcommand{\changeTwo}[1]{\textcolor{black}{#1}}

\begin{document}
\title{Adaptive Robust Data-driven Building Control via Bi-level Reformulation: an Experimental Result}
\author{Yingzhao Lian$^\dagger$, Jicheng Shi$^\dagger$, Manuel Koch and Colin N. Jones$^*$
\thanks{This work was supported by the Swiss National Science Foundation under the RISK project (Risk Aware Data-Driven Demand Response), grant number 200021 175627}
\thanks{$^\dagger$: First two authors contribute equally. In particular, Yingzhao Lian is responsible for the framework of this work and Jicheng Shi is responsible for the experiment.}
\thanks{*: Corresponding author}
\thanks{Yingzhao Lian, Jicheng Shi, Manuel Koch and Colin Neil Jones are with the Automatic Control Lab, EPFL, Switzerland.(e-mail: \{yingzhao.lian, jicheng.shi, manuelpascal.koch, colin.jones\}@epfl.ch)}\vspace{-1em}
}
\maketitle

\begin{abstract}
Data-driven control approaches for the minimization of energy consumption of buildings have the potential to significantly reduce deployment costs and increase uptake of advanced control in this sector. A number of recent approaches based on the application of Willems' fundamental lemma for data-driven controller design from input/output measurements are very promising for deterministic LTI systems. This paper \change{proposes a systematic way to handle unknown measurement noise and measurable process noise}, and extends these data-driven control schemes to adaptive building control via a robust bi-level formulation, whose upper level ensures robustness and whose lower level guarantees prediction quality. Corresponding numerical improvements and an active excitation mechanism are proposed to enable a computationally efficient reliable operation. The efficacy of the proposed scheme is validated by \change{a multi-zone building simulation} and a real-world experiment on a single-zone conference building on the EPFL campus. The real-world experiment includes a 20-day non-stop test, where, without extra modeling effort, our proposed controller improves 18.4\% energy efficiency against an industry-standard controller, while also robustly ensuring occupant comfort.
\end{abstract}


\section{Introduction}
Predictive control techniques have recently been proven to be effective for controlling building \change{heating, ventilation, and air conditioning (HVAC)} systems with reduced energy consumption while simultaneously improving comfort. However, the cost of model building and commissioning has proven to be very high and as a result, data-driven approaches for building energy saving have been receiving broad attention.

The motivation of this work is to develop a practical robust adaptive data-driven building controller, whose time-varying linear dynamics are disturbed by measurable process noise including outdoor temperature and solar radiation, and whose output measurements are contaminated by unknown measurement noise. Beyond building control, systems with measurable disturbances are ubiquitous, especially in energy-related applications: solar radiation in photovoltaic power systems, electricity demand in power grids, and power generation in airborne wind energy systems, to name a few. The main contributions of this paper are summarized as follows:
\begin{itemize}
    \item \change{Propose a novel trajectory prediction method, where a new upper bound of the Wasserstein distance between the trajectory determined by the Willems' fundamental lemma and the noisy measured sequence is minimized.}
    \item Propose a tractable robust bi-level data-driven building controller,  \change{whose lower level problem applies the trajectory prediction scheme proposed in the last point.}
    \item \change{Propose a numerically efficient extension of the proposed bi-level scheme to linear time-varying dynamics. This includes an active excitation scheme to maintain robust online data updates and numerical details for efficient \changeTwo{online update of the matrix inversion used in the algorithm.}}
    \item Experimental validation of the proposed scheme on a real-world building.
\end{itemize}

\subsection*{Previous Work}
Data can be used to model building dynamics~\cite{boodi2018intelligent} or to directly generate/improve control policies. Due to seasonal variations~\cite{afroz2018modeling} and component wear, building dynamics are usually slowly time-varying, and adaptive model predictive control (MPC) has been introduced to combine online parameter estimation and control in~\cite{drgovna2020all}. For example, the experiment in~\cite{tesfay2018adaptive} adaptively \change{estimates} the model of an evaporator in the HVAC system, which \change{is} then used to control the valve set-points. \cite{maasoumy2013online} \change{runs} an extended Kalman filter to achieve online parameter adaptation before the estimated model \change{is} used in an MPC controller. However, these parameter-estimation-based adaptive methods usually require a-priori knowledge about the structure of the building dynamics and/or the HVAC systems.

Beyond running through a modeling/estimation procedure, data can be used to refine a control policy. The main approaches in this direction include reinforcement learning (RL)~\cite{sutton2018reinforcement} and iterative learning control (ILC)~\cite{bristow2006survey}. In particular, ILC has been used for buildings with fixed heating/occupant schedules~\cite{minakais2014groundhog,yan2010iterative} and RL for learning a building control policy that is not necessarily iterative~\changeTwo{\cite{ceusters2021model}}. To run these learning schemes, a high-fidelity building simulator is usually required, and therefore publications on successful experiments with HVAC systems are rare, with a few exceptions being~\cite{mocanu2018line,costanzo2016experimental}\changeTwo{,~\cite{zhang2018practical, kazmi2016generalizable, ruelens2016reinforcement} }. 

Beyond RL and ILC, data can also be used to directly characterize the system's responses from a behavioral theoretic viewpoint~\cite{willems1997introduction}. Willems' fundamental lemma is such a tool that provides a characterization of linear time invariant (LTI) systems from measured input and output trajectories. Such a characterization offers a convenient interface to data-driven controller design~\cite{markovsky2007linear,markovsky2008data,coulson2019data,de2019formulas}. Motivated by the simplicity and effectiveness of the fundamental lemma, its extension to a more general setting is attracting broad attention, including nonlinear extensions~\cite{berberich2020trajectory,bisoffi2020data,rueda2020data,lian2021nonlinear}\changeTwo{, data informativity~\cite{yu2021controllability}, discriptor system~\cite{schmitz2022willems}, e.t.c~\cite{van2022data}.}

\change{Even for LTI systems, the absence of output measurement noise in the standard Willems' fundamental lemma limits its practicality.  To accommodate this issue, a wide range of researchers are studying this challenge. In~\cite{berberich2020robust,van2021matrix,berberich2020robust2}, classic robust control design tools such as linear fractional transforms have been proposed to design robust linear feedback controllers. Parallel to the studies in linear feedback control laws, robustness in predictive control schemes has also been studied \cite{coulson2019regularized, dorfler2022bridging,xu2021data}, where regularization is the main tool applied to deal with measurement noise. \cite{coulson2019regularized} shows that the regularization is related to the distributional robustness of the system uncertainty (including measurement noise). \cite{dorfler2022bridging} studies a different viewpoint, where the regularization is linked to the loss function used in the system identification.  When convex relaxation is applied, \changeTwo{the} loss function in the system identification procedure results in a regularization term~\cite{dorfler2022bridging,markovsky2021behavioral}.}

\change{Building on this regularization viewpoint, data-driven controllers based on the Willems' fundamental lemma have been successfully deployed on different real-world systems, such as a quadrotor~\cite{elokda2021data}, a four-tank system~\cite{berberich2021data}, etc. However, tuning the regularization weight still poses a non-trivial challenge, and an exhaustive search based on simulation is commonly used~\cite{markovsky2021behavioral}.}

\change{Motivated by the discussions above, this paper proposes a novel bi-level data-driven  predictive controller. Instead of solving the general case with unknown process/measurement noise, this paper focuses on applications with unknown measurement noise and measurable process noise; such as building control (see Section.~\ref{sect:stage} for more details). In the following, background knowledge is reviewed in Section~\ref{sect:pre}. In Section~\ref{sect:stage}, the problem setup is presented and the proposed predictive controller is summarized in~\eqref{eqn:rb_deepc_og}. Section~\ref{sect:LTI_deepc} and Section~\ref{sect:LTV_deepc} then accordingly develop and analyze the proposed scheme in the linear time-invariant (LTI) and the linear time-varying (LTV) cases accordingly. Regarding the LTV systems, an active excitation mechanism is introduced in Section~\ref{sect:act} accompanied by numerical details of efficient online updates in Section~\ref{sect:numerical}. The complete robust adaptive data-driven controller is summarized in Algorithm~\ref{alg:active_deepc}. The effectiveness of the proposed algorithm is first validated through temperature control of a multi-zone building in Section~\ref{sect:simulation}, followed by the experimental results with an entire building on the EFPL campus in Section~\ref{sect:experiment}. A conclusion of this paper is given in Section~\ref{sect:conclusion}.}

\vspace{0.5em}

\noindent\textbf{Notation:} $\mathcal{N}(\mu,\Sigma)$ denotes a Gaussian distribution with mean $\mu$ and covariance $\Sigma$. $\otimes$ is the symbol for the Kronecker product. $\times$ is the symbol for the set product operation \change{(i.e. $A\times B:= \{(x,y)|x\in A,\;y\in B\}$)}, $\oplus$ is the symbol of Minkowski sum \change{(i.e. $A\oplus B:=\{x|\exists\;x_1\in A,\;x_2\in B,\; x=x_1+x_2\}$)} and $\ominus$ is the symbol of \change{Pontryagin} difference \change{(i.e. $A\ominus B :=\{x|x+y\in A,\;\forall\;y\in B\}$)}. $\text{colspan}(A)$ denotes the column space (i.e. range) of the matrix A. $I_n$ is the $n\times n$ identity matrix, \change{$\textbf{O}$ and $\mathbf{0}$ are zero matrix and zero vector respectively}. $\lVert \cdot\rVert$ denotes the Euclidean two-norm. $x: = \{x_i\}_{i=1}^T$ denotes a sequence of size $T$ indexed by $i$. $x_i$ denotes the measurement of $x$ at time $i$, and $x_{1:L}:=[x_1^\top,x_2^\top\dots x_L^\top]^\top$ denotes a concatenated sequence of $x_i$ ranging from $x_1$ to $x_L$, and we drop the index to improve clarity if the intention is clear from the context.

\section{Preliminaries}\label{sect:pre}
\begin{definition}
A Hankel matrix of depth $L$ associated with a vector-valued signal sequence $s:=\{s_i\}_{i=1}^T,\;s_i\in\R^{n_s}$ is
\begin{align*}
    \Han_L(s):=
    \begin{bmatrix}
    s_1 & s_2&\dots&s_{T-L+1}\\
    s_2 & s_3&\dots&s_{T-L+2}\\
    \vdots &\vdots&&\vdots\\
    s_{L} & s_{L+1}&\dots&s_T
    \end{bmatrix}\;.
\end{align*}
\end{definition}
\vspace{1em}

A deterministic linear time-invariant (LTI) system, dubbed $\mathfrak{B}(A,B,C,D)$, is defined as
\begin{align}\label{eqn:lin_dyn_deter}
\begin{split}
    x_{i+1} = Ax_i+Bu_i\;,\;y_i = Cx_i+Du_i\;,
\end{split}
\end{align}
whose order is $n_x$. $n_u,\enspace n_y$ denote its input and output dimensions respectively. An $L$-step trajectory generated by this system is 
\begin{align*}
    \begin{bmatrix}u_{1:L} & y_{1:L}\end{bmatrix}:=\begin{bmatrix}
         u_1^\top&\dots & u_{L}^\top & y_{1}^\top &\dots&y_{L}^\top
    \end{bmatrix}^\top\;.
\end{align*} 
The set of all possible $L$-step trajectories generated by $\mathfrak{B}(A,B,C,D)$ is denoted by $\mathfrak{B}_L(A,B,C,D)$. 

\change{For the sake of consistency, a datapoint coming from the historical dataset is marked by boldface subscript $_\textbf{d}$.} Given a sequence of input-output measurements $\{u_{\textbf{d},i},y_{\textbf{d},i}\}_{i=1}^T$, we call the input sequence persistently exciting of order $L$ if $\Han_L(u_\textbf{d})$ is full row rank. By building the following $n_c$-column stacked Hankel matrix 
\begin{align}\label{eqn:fund_column}
    \Han_L(u_\textbf{d}, y_\textbf{d}):=\begin{bmatrix}
    \Han_{L}(u_{\textbf{d}})\\\Han_L(y_{\textbf{d}})
    \end{bmatrix}\;,
\end{align}
we state \textbf{Willems' Fundamental Lemma} as
\begin{lemma}\label{lem:funda}\cite[Theorem 1]{willems2005note}
Consider a controllable linear system and assume $\{u_\textbf{d}\}_{i=1}^T$ is persistently exciting of order $L+ n_x$. The condition $\text{colspan}(\Han_L(u_\textbf{d},y_\textbf{d}))=\mathfrak{B}_L(A,B,C,D)$ holds.
\end{lemma}

For the sake of consistency, $L$ is reserved for the length of the system responses and $n_c$ denotes the number of columns in a Hankel matrix. 

A data-driven control scheme has been proposed in~\cite{coulson2019data,markovsky2007linear}, where Lemma~\ref{lem:funda} generates a trajectory prediction. \change{We postpone the detailed discussion about the existing methods to Section~\ref{sect:compare}, in order to better articulate the idea behind the existing methods and to better compare the difference between the proposed and the existing methods.}

\subsection{Wasserstein Distance}\label{sect:wass_dist}
The $2$-Wasserstein distance between two distributions $\mathbb{P}_x$ and $\mathbb{P}_y$ is defined by
\begin{align*}
    W(\mathbb{P}_x,\mathbb{P}_y):= \left(\inf\limits_{\gamma\in\Gamma(\mathbb{P}_x,\mathbb{P}_y)}\E_{(x,y)\sim\gamma}\lVert x-y\rVert^2\right)^\frac{1}{2}\;,
\end{align*}
where $\Gamma(\mathbb{P}_x,\mathbb{P}_y)$ is the family of joint distributions whose marginals are $\mathbb{P}_x$ and $\mathbb{P}_y$. The 2-Wasserstein distance models the optimal transport between $\mathbb{P}_x$ and $\mathbb{P}_y$ in terms of the Euclidean distance. If $\mathbb{P}_x\sim\mathcal{N}(\mu_x,\Sigma_x)$ and $\mathbb{P}_y\sim\mathcal{N}(\mu_y,\Sigma_y)$, then the squared 2-Wasserstein distance has a closed-form~\cite{dowson1982frechet}
\begingroup\makeatletter\def\f@size{8.5}\check@mathfonts
\begin{align}\label{eqn:wass_gauss}
    W(\mathbb{P}_x,\mathbb{P}_y)^2 = \lVert\mu_x-\mu_y\rVert^2+\tr\left(\Sigma_x+\Sigma_y-2(\Sigma_x^{\frac{1}{2}}\Sigma_y\Sigma_x^{\frac{1}{2}})^\frac{1}{2}\right)
\end{align}
\endgroup

\subsection{Setting the Stage}\label{sect:stage}
\change{Recall the main properties in building applications:
\begin{itemize}
    \item The system dynamics are slowly time-varying.
    \item The building dynamics evolve under strong \textbf{measurable} process noise, such as solar radiation and outdoor temperature. 
    \item Because of the activity of the occupants, the output measurements, particularly the measurements of the indoor temperature, are noisy.
\end{itemize}
Given these properties, this work focuses on the following uncertain linear time-varying (LTV) system}
\begin{align}\label{eqn:dyn_varying}
\begin{split}
    x_{i+1} &= A_ix_i + B_iu_i+ E_i w_i\\
    \overline{y}_i &= C_ix_i + D_iu_i\;,\;y_i = \overline{y}_i+ v_i
\end{split}
\end{align}
where $w_i\in\R^{n_w}$ is bounded measurable process noise and $v_i\sim\mathcal{N}(0,\Sigma_v)$ with $v_i\in\R^{n_y}$ independent and identically distributed (i.i.d) unknown measurement noise. In particular, $\overline{y}$ is the system output, which is unknown, and $y$ is the measurement read out from the sensors. In particular, in the building control problem, $w$ mostly reflects the external temperature, solar radiation, occupancy, etc. \change{Note that LTI systems are special case of the LTV systems~\eqref{eqn:dyn_varying}.}

In a similar manner to~\cite{huang2019decentralized,lian2021system}, by viewing $w$ as uncontrolled inputs, Lemma~\ref{lem:funda} can be generalized and the corresponding $L$-step trajectory is augmented to
\begin{align*}
    [u_{1:L}\enspace w_{1:L}\enspace y_{1:L}]:=[u_1^\top \;\dots&u_{L}^\top\;w_1^\top&\dots\;w_{L}^\top\;y_{1}^\top&\dots\;y_{L}^\top]^\top.
\end{align*}

\change{
Based on this idea, the rest of this paper will develop and analyze the following bi-level predictive control problem step by step:
\begin{subequations}\label{eqn:rb_deepc_og}
\begin{align}
        \min\limits_{\substack{y_{pred}\\\overline{u}_{pred},K}} \;&\; J(y_{pred},u_{pred})\nonumber\\
        \text{s.t.} \;&\forall\;\tilde{w}_{pred} \in \overline{w} \oplus \tilde{\mathcal{W}}\label{eqn:rb_deepc_og_wcons}\\
        &\; u_{pred}= \overline{u}_{pred}+K\tilde{w}_{pred}\in \tilde{\mathcal{U}}\;,\label{eqn:rb_deepc_og_ucons}\\
        &\;y_{pred} = \Han_{L,pred}(y_\textbf{d})g \in \mathcal{Y}\;,\label{eqn:rb_deepc_og_ycons}\\
        &g\in\argmin_{g_l,\sigma_l} \frac{1}{2}\lVert\sigma_l\rVert^2+\frac{1}{2}g_l^\top\mathcal{E}_g g_l\label{eqn:rb_deepc_og_lower}\\
        &\quad\quad\quad\text{s.t.}\;\begin{bmatrix}
        \Han_{L,init}(y_\textbf{d})\\\Han_{L,init}(u_\textbf{d})\\\Han_{L,init}(\tilde{w}_\textbf{d})\\\Han_{L,pred}(u_\textbf{d})\\\Han_{L,pred}(\tilde{w}_\textbf{d})
        \end{bmatrix}g_l=\begin{bmatrix}
        y_{init}+\sigma_l\\u_{init}\\\tilde{w}_{init}\\u_{pred}\\\tilde{w}_{pred}
    \end{bmatrix}\label{eqn:rb_deepc_og_lower_cons}\;,
\end{align}
\end{subequations}
where $J(\cdot,\cdot)$ is a convex objective function, $\tilde{\mathcal{U}}$ and $\mathcal{Y}$ denote compact convex constraint sets on the input and the output. $\overline{w}\oplus\tilde{\mathcal{W}}$ models the set of possible future process noise with $\overline{w}$ the nominal future process noise and $\tilde{\mathcal{W}}$ the quantified uncertainty. Additionally, the tight input constraint is $\mathcal{U}$, and $\tilde{\mathcal{U}}$ is its tightening (i.e. $\tilde{\mathcal{U}}\subseteq\mathcal{U}$). $\tilde{\mathcal{W}}$ is an augmented set of actual process noise, and the set of the actual process noise is $\mathcal{W}$ (i.e. $\exists\;n\geq 0$, such that $\tilde{\mathcal{W}}\subset \mathcal{W}\times\R^n$). The selection of both $\tilde{\mathcal{U}}$ and $\tilde{\mathcal{W}}$ will be elaborated in the following sections.  $u_{init},\tilde{w}_{init}, y_{init}$ are $t_{init}$-step sequences of the measured inputs, process noise and outputs preceding the current point in time. Accordingly, $u_{pred},\tilde{w}_{pred},y_{pred}$ are the corresponding $n_h$-step predictive sequences viewed from the current time step. The matrix $\Han_L(y_\textbf{d})$ is split into two sub-Hankel matrices:
\begin{align*}
    \Han_L(y_\textbf{d}) = \begin{bmatrix}
        \Han_{L,init}(y_\textbf{d})\\\Han_{L,pred}(y_\textbf{d})
    \end{bmatrix}\;.
\end{align*}
The matrix $\Han_{L,init}(y_\textbf{d})$ is of depth $t_{init}$ and the depth of $\Han_{L,pred}(y_\textbf{d})$ is the prediction horizon $n_h$ such that $t_{init}+n_h = L$. The matrices $\Han_{L,init}(u_\textbf{d})$, $\Han_{L,pred}(u_\textbf{d})$, $\Han_{L,init}(\tilde{w}_\textbf{d})$, $\Han_{L,pred}(\tilde{w}_\textbf{d})$ are defined similarly. The choice of $t_{init}$ is made to ensure a unique estimation of the initial state; please refer to~\cite{markovsky2008data} for more details. Further, recall the condition of Willems' fundamental lemma~\ref{lem:funda}, the proposed scheme requires the following assumption:}
\begin{assumption}\label{ass:PE}
$\{u_{\mathbf{d}},\tilde{w}_{\mathbf{d}}\}$ is \changeTwo{persistently exciting} of order $L+n_x$.
\end{assumption}

\change{
The proposed scheme~\eqref{eqn:rb_deepc_og} is bi-level, where the penalty weight $\mathcal{E}_g$, the set of disturbances $\tilde{\mathcal{W}}$, and the tightened input constraints $\tilde{\mathcal{U}}$ vary with respect to the working conditions. The selections of these terms will be elaborated in the following sections. In Section~\ref{sect:LTI_deepc}, a rigorous setup for LTI systems will first be developed to show the logic behind the bi-level formulation. The extension to LTV system and the computational details will then be discussed in Section~\ref{sect:LTV_deepc}.}

\section{Robust Bi-level Data-driven Control for LTI Systems}\label{sect:LTI_deepc}
In this part, we will establish a rigorous mathematical framework for our proposed bi-level scheme~\eqref{eqn:rb_deepc_og} by considering the LTI version of the targeted time-varying dynamics~\eqref{eqn:dyn_varying}. The LTI dynamics satisfy:
\change{
\begin{align*}
A_i=A_j,\;B_i=B_j,\;C_i=C_j,\;D_i=D_j,\;E_i=E_j,\forall\;i,j\;.
\end{align*}}
\change{Note that the output trajectory prediction with respect to a given control input sequence $u_{pred}$ is the key component for predictive control. In this section, we will show that this can be done by minimizing a Wasserstein distance upper bound between the trajectories determined by the fundamental lemma and noisy measurement sequences (Lemma~\ref{lem:wasser_bound}). This trajectory prediction problem defines the lower level problem in our proposed scheme~\eqref{eqn:rb_deepc_og}, whose single-level reformulation is summarized in  Lemma~\ref{lem:rb_deepc}. In the end, Subsection~\ref{sect:compare} concludes this Section~\ref{sect:LTI_deepc} by comparing the proposed scheme with the existing single-level schemes.}

\subsection{Wasserstein Prediction Upper Bound}\label{sect:wasser_bound}
Regarding~\eqref{eqn:dyn_varying}, the system output $\overline{y}_{init}$ is contaminated by measurement noise, giving a noisy measurement vector $y_{init}$. The output $\overline{y}_{init}$ thus follows the distribution (Section~\ref{sect:stage}):
\begin{align*}
    \overline{y}_{init}\sim\mathcal{N}(y_{init},\Sigma_{init})\;,
\end{align*}
where $\Sigma_{init}=I_{t_{init}}\otimes \Sigma_v$. Similarly, the measurement Hankel matrix $\Han_{L,init}(y_\textbf{d})$ is subject to measurement noise, and we denote the uncertain system output Hankel matrix by $\Han_{L,init}(\overline{y_\textbf{d}})$. Then, for arbitrary $g\in\R^{n_c}$, the following lemma quantifies the distribution distance between an uncertain trajectory generated by the fundamental lemma, $\Han_{L,init}(\overline{y}_\textbf{d})g$, and the uncertain system output sequence $\overline{y}_{init}$

\begin{lemma}\label{lem:wasser_bound}
$\forall\; g\in\R^{n_c}$, the squared Wasserstein distance $W(\Han_{L,init}\left(\overline{y_\textbf{d}}\right)g,\overline{y}_{init})^2$ is upper bounded by 
\[\lVert\Han_{L,init}(y_\textbf{d})g-y_{init}\rVert^2+\left(\sqrt{t_{init}}\lVert g\rVert-1\right)^2\tr(\Sigma_v)\;.\]
\end{lemma}

\begin{proof}
Similar to the distribution of $\overline{y}_{init}$, the distribution of the $i$-th column of the output Hankel matrix $\Han_{L,init}(\overline{y})$ follows a Gaussian distribution $\mathcal{N}(y_{\textbf{d},i:i+t_{init-1}},\Sigma_{init})$ and the adjacent columns are correlated. By the basic properties of Gaussian distributions, we have
\begin{align*}
    \Han_{L,init}(\overline{y_\textbf{d}})g\sim \mathcal{N}(\Han_{L,init}(y_\textbf{d})g,\tilde{\Sigma}_{init})\;,
\end{align*}
where $\tilde{\Sigma}_{init}=\tilde{G}\otimes\Sigma_v$ and the entry of $\tilde{G}\in\R^{t_{init}\times t_{init}}$ is
\begin{align*}
    \tilde{G}_{i,j} = \sum\limits_{n=1}^{n_c-|i-j|} g_ng_{n+|i-j|}\;.
\end{align*}

Hence, the squared Wasserstein distance $W(\Han_{L,init}(\overline{y_\textbf{d}})g,\overline{y}_{init})^2$ is
\begin{align*}
W(\Han_{L,init}(\overline{y_\textbf{d}})g,&\overline{y}_{init})^2=\lVert\Han_{L,init}(y_\textbf{d})g-y_{init}\rVert^2\\&+ \underbrace{\tr(\tilde{\Sigma}_{init}+\Sigma_{init}-2(\Sigma_{init}^\frac{1}{2}\tilde{\Sigma}_{init}\Sigma_{init}^\frac{1}{2})^\frac{1}{2})}_{(a)}\;.
\end{align*}
In the rest of this proof, we will upper bound the term $(a)$ above,
\begin{align}\label{eqn:prf_lem_wasser}
\begin{split}
        (a)&=\tr(\tilde{\Sigma}_{init})+\tr(\Sigma_{init})+2\tr\left((\Sigma_{init}^\frac{1}{2}\tilde{\Sigma}_{init}\Sigma_{init}^\frac{1}{2})^\frac{1}{2}\right)\\
    &=\left(\tr(\tilde{G})+\tr(I_{t_{init}})\right)\tr(\Sigma_v)-2\underbrace{\tr\left(\left((\tilde{G}I_{t_{init}})\otimes\Sigma_v^2\right)^\frac{1}{2}\right)}_{(b)}\\
    &=t_{init}(\lVert g\rVert^2+1)\tr(\Sigma_v)-2\underbrace{\tr(\tilde{G}^\frac{1}{2}\otimes\Sigma_v)}_{(c)}\\
    &=\left(t_{init}(\lVert g\rVert^2+1)-2\tr(\tilde{G}^\frac{1}{2})\right)\tr(\Sigma_v)\\
    &\stackrel{(d)}{\leq}\left(t_{init}(\lVert g\rVert^2+1)-2\tr(\tilde{G})^\frac{1}{2}\right)\tr(\Sigma_v)\\
    &= \left(t_{init}(\lVert g\rVert^2+1)-2\sqrt{t_{init}}\lVert g\rVert\right)\tr(\Sigma_v)\\
    &\stackrel{(e)}{\leq}\left(\sqrt{t_{init}}\lVert g\rVert-1\right)^2\tr(\Sigma_v)\;,
\end{split}
\end{align}
where terms $(b)$, $(c)$ follow the same fact that $(A\otimes B)(C\otimes D)=AC\otimes BD$. The inequality $(d)$ follows~\cite[Theorem 1 (ii)]{neudecker1992matrix}
\begin{align*}
    \tr(\tilde{G})^\frac{1}{2}\leq\tr(\tilde{G}^\frac{1}{2})\;.
\end{align*}
The inequality $(e)$ uses the fact that $t_{init}\geq 1$. We conclude the proof with
\begin{align*}
    W(\Han_{L,init}&\left(\overline{y_\textbf{d}}\right)g,\overline{y}_{init})^2\leq\\
    &\lVert\Han_{L,init}(y_\textbf{d})g-y_{init}\rVert^2+\left(\sqrt{t_{init}}\lVert g\rVert-1\right)^2\tr(\Sigma_v)\;.\;
\end{align*}
\end{proof}

Based on the upper bound in Lemma~\ref{lem:wasser_bound}, we can generate a trajectory prediction via the following optimization problem: 
\begin{subequations}\label{eqn:pred_wasser}
\begin{align}
    y_{pred}=& \Han_{L,pred}(y_\textbf{d})g(\tilde{w}_{pred})\label{eqn:E_pred}\\
    g(\tilde{w}_{pred})\in&\argmin_{g_l,\sigma_l} \frac{1}{2}\lVert\sigma_l\rVert^2+\frac{1}{2}\left(\sqrt{t_{init}}\lVert g_l\rVert-1\right)^2\tr(\Sigma_v)\notag\\
        &\quad\text{s.t.}\;\begin{bmatrix}
        \Han_{L,init}(y_\textbf{d})\\\Han_{L,init}(u_\textbf{d})\\\Han_{L,init}(\tilde{w}_\textbf{d})\\\Han_{L,pred}(u_\textbf{d})\\\Han_{L,pred}(\tilde{w}_\textbf{d})
        \end{bmatrix}g_l=\begin{bmatrix}
        y_{init}+\sigma_l\\u_{init}\\\tilde{w}_{init}\\u_{pred}\\\tilde{w}_{pred}
        \end{bmatrix}\;.\label{eqn:pred_eq}
\end{align}
\end{subequations}
\change{Equation~\eqref{eqn:pred_eq} gives the trajectory expectation generated by the Willems' fundamental lemma. Meanwhile, $\begin{bmatrix}y_{init}^\top \changeTwo{+\sigma_l^\top} &u_{init}^\top & \tilde{w}_{init}^\top& u_{pred}^\top & \tilde{w}_{pred}^\top\end{bmatrix}^\top$ is the trajectory composed by noisy measurements read out from the sensors, the planned future input and a predicted disturbance trajectory.} Thus, optimization problem~\eqref{eqn:pred_wasser} minimizes the discrepancy upper bound between an I/O sequence estimated by measurement and an I/O trajectory estimated by the fundamental lemma.
\begin{remark}\label{rmk:pred_uncertain}
Note that the Hankel matrix $\Han_{L,pred}(y_\textbf{d})$ is also subject to measurement noise ($v$ in~\eqref{eqn:dyn_varying}), which means that the estimate of a fundamental lemma based predictive trajectory $\Han_{L,pred}(\overline{y}_\textbf{d})g$ is also uncertain and Gaussian with expectation $\Han_{L,pred}(y_\textbf{d})g$. The prediction given by~\eqref{eqn:E_pred} can therefore be considered as a certainty equivalent prediction. Meanwhile, even though it is possible to consider the uncertainty in $y_{pred}$, for the sake of a clean layout, we only use the certainty equivalence prediction~\eqref{eqn:pred_wasser} in this paper. \changeTwo{It is noteworthy to mention that, based on our experience, the quantified uncertainty of $w_{pred}$ offered by the weather forecast is usually highly conservative. Using certainty equivalence is empirically sufficient to give desirable robust control performance in building applications (see Section~III-B for more details).}
\end{remark}

\subsection{Tractable Bi-level Reformulation}\label{sect:bi_reform}
\change{In this section, we will develop the proposed scheme~\eqref{eqn:rb_deepc_og} by integrating the prediction problem~\eqref{eqn:pred_wasser} into a predictive control problem.} This controller should maintain the system performance while ensuring robust constraint satisfaction regardless of future realizations of the process noise, and so we assume that the future process noise $\tilde{w}_{pred}$ can be predicted with uncertainty quantification. Buildings are systems satisfying this assumption. For example, the weather forecast can provide a future temperature prediction with an uncertainty tube centered around a nominal prediction, such that the actual future temperature realization fluctuates within this tube. Therefore, we denote the set of predicted future process noise realizations by $\tilde{w}_{pred} \in \overline{w}\oplus\tilde{\mathcal{W}}$, where $\overline{w}$ is the nominal prediction and $\tilde{\mathcal{W}}$ denotes the uncertainty tube. And we state our assumption:
\begin{assumption}\label{ass:process_noise}
$\overline{w}$ and $\tilde{\mathcal{W}}$ are known.
\end{assumption}

To make the controller less conservative, we consider a predictive control input with a linear feedback from process noise
\begin{align}\label{eqn:u_fb}
    u_{pred} = \overline{u}_{pred}+K\tilde{w}_{pred}\;,
\end{align}
where the feedback control $K$ is a decision variable in our predictive control problem. In particular, $\overline{u}$ reflects the nominal control inputs while feedback $K$ adapts the control input based on the actual realization of the process noise.

\change{By replacing the lower level problem in~\eqref{eqn:rb_deepc_og} (i.e. Equation~\eqref{eqn:rb_deepc_og_lower} and~\eqref{eqn:rb_deepc_og_ycons}) with the prediction problem~\eqref{eqn:pred_wasser}, we can state a bi-level robust predictive control problem for LTI systems:}
\begin{subequations}\label{eqn:rb_deepc_LTI_ideal}
\begin{align} 
        \min\limits_{\substack{\overline{y}_{pred}\\\overline{u}_{pred},K}} \;&\; J(y_{pred},u_{pred})\\
        \text{s.t.} \;&\forall\;\tilde{w}_{pred} \in \overline{w}\oplus\tilde{\mathcal{W}}\nonumber\\
        &\; u_{pred}= \overline{u}_{pred}+K\tilde{w}_{pred}\in \tilde{\mathcal{U}}\;,\nonumber\\
        &\;y_{pred} = \Han_{L,pred}(y_\textbf{d})g(\tilde{w}_{pred}) \in \mathcal{Y}\;,\label{eqn:rb_deepc_LTI_ideal_ycons}\\
        g(\tilde{w}_{pred})&\in\argmin_{g_l,\sigma_l} \frac{1}{2}\lVert\sigma_l\rVert^2+\frac{1}{2}\left(\sqrt{t_{init}}\lVert g_l\rVert-1\right)^2\tr(\Sigma_v)\nonumber\\
        &\quad\quad\quad\text{s.t.}\;\begin{bmatrix}
        \Han_{L,init}(y_\textbf{d})\\\Han_{L,init}(u_\textbf{d})\\\Han_{L,init}(\tilde{w}_\textbf{d})\\\Han_{L,pred}(u_\textbf{d})\\\Han_{L,pred}(\tilde{w}_\textbf{d})
        \end{bmatrix}g_l=\begin{bmatrix}
        y_{init}+\sigma_l\\u_{init}\\\tilde{w}_{init}\\u_{pred}\\\tilde{w}_{pred}
    \end{bmatrix}\nonumber\;,
\end{align}
\end{subequations}
\change{where we set $\tilde{\mathcal{W}}=\mathcal{W}$ and $\tilde{\mathcal{U}}=\mathcal{U}$. Problem~\eqref{eqn:rb_deepc_LTI_ideal} is a bi-level optimization problem, whose upper level decides the optimal control and whose lower level generates the corresponding predictive output trajectory. In particular, the upper level problem sends $\overline{u}_{pred}$ and $K$ to the lower level problem. Then for each specific $\tilde{w}_{pred}$, the lower level problem returns the corresponding predictor $g(\tilde{w}_{pred})$, which accordingly defines the output trajectory predictions $y_{pred}$ for that $\tilde{w}_{pred}$ (Equation~\eqref{eqn:rb_deepc_LTI_ideal_ycons}).} In turn, the robust constraint in the upper level ensures that input and output constraints are satisfied for all $\tilde{w}_{pred}$ in the considered set $\overline{w}\oplus\mathcal{W}$. In conclusion, the upper level problem optimizes $\overline{u}_{pred}$ and $K$ based on the prediction given by the lower level problem. 

However, this bi-level problem~\eqref{eqn:rb_deepc_LTI_ideal} is hard to solve numerically, because the objective in the lower level problem is non-convex.\footnote{To ensure a composition of two convex functions is convex, the outer convex function needs be non-decreasing~\cite[Chapter 3.2.4]{boyd2004convex}. To see the non-convexity in~\eqref{eqn:rb_deepc_LTI_ideal}, one can plot the following function $f(x) = (|x|-1)^2,x\in\R$.} To address this issue, we state a looser, but convex, Wasserstein upper bound in the following corollary.
\begin{corollary}\label{cor:wasser_bound}
$\forall\; g\in\R^{n_c}$, the squared Wasserstein distance $W(\Han_{L,init}\left(\overline{y_\textbf{d}}\right)g,\overline{y}_{init})^2$ is upper bounded by
\begin{align*}\lVert\Han_{L,init}(y_\textbf{d})g-y_{init}\rVert^2+t_{init}\tr(\Sigma_v)\left(\lVert g\rVert^2+1\right).\end{align*}
\end{corollary}
\begin{proof}
In the inequality $(d)$ of Equation~\eqref{eqn:prf_lem_wasser}, we have $\tr(\tilde{G})\geq0$ as $\tilde{G}$ is positive semi-definite. Therefore, we give a convex Wasserstein distance upper bound as
\begin{align*}
    &W\left(\Han_{L,init}\left(\overline{y_\textbf{d}}\right)g,\overline{y}_{init}\right)^2\nonumber\\
    &\quad\leq\lVert\Han_{L,init}(y_\textbf{d})g-y_{init}\rVert^2+t_{init}^2\tr(\Sigma_v)\left(\lVert g\rVert^2+1\right)\;.
\end{align*}
\end{proof}

Bringing everything together, \change{we eliminate the solution of the nonconvex problem~\eqref{eqn:rb_deepc_LTI_ideal}, and arrive at a convex tractable version (i.e. the proposed scheme~\eqref{eqn:rb_deepc_og}). This numerically tractable predictive control problem~\eqref{eqn:rb_deepc_og} will be used throughout the paper. The terms $\tilde{\mathcal{U}},\;\tilde{\mathcal{W}}$ and $\mathcal{E}_g$ used in the LTI case are:
\begin{itemize}
    \item $\mathcal{E}_g = t_{init}^2\tr(\Sigma_v)I_{n_c}$:  The lower level problem minimizes the convex Wasserstein upper bound given in Corollary~\ref{cor:wasser_bound}. For the sake of clarity, the constant term $t_{init}^2\tr(\Sigma_v)$ is dropped in its objective.
    \item $\tilde{\mathcal{W}}=\mathcal{W},\;\tilde{\mathcal{U}}=\mathcal{U}$: The set of predictive process noise is directly defined by the process noise forecast ( Assumption~\ref{ass:process_noise}), and the input constraint is not tightened.
\end{itemize}}

\change{Up to this point, we have derived the bi-level structure used in the proposed scheme~\eqref{eqn:rb_deepc_og}, and we will show that it has a tractable single-level reformulation via the following lemma:}
\begin{lemma}\label{lem:rb_deepc}
The following single-level robust optimization problem is equivalent to the bi-level \change{problem~\eqref{eqn:rb_deepc_og}}. 
\begin{subequations}\label{eqn:rb_deepc}
\begin{align}
        \min\limits_{\substack{\overline{y}_{pred}\\\overline{u}_{pred},K}} \;&\; J(y_{pred},u_{pred})\nonumber\\
        \text{s.t.} \;&\forall\;\tilde{w}_{pred} \in \overline{w}\oplus\tilde{\mathcal{W}}\nonumber\\
        &\; u_{pred}= \overline{u}_{pred}+K\tilde{w}_{pred}\in \tilde{\mathcal{U}}\;,\nonumber\\
        &\;y_{pred} = \Han_{L,pred}(y_\textbf{d})g(\tilde{w}_{pred}) \in \mathcal{Y}\;,\nonumber\\
        & \;\begin{bmatrix}
        g(\tilde{w}_{pred})\\\kappa
        \end{bmatrix} = M^{-1}\begin{bmatrix}\Han_{L,init}(y_\textbf{d})^\top y_{init}\\u_{init}\\\tilde{w}_{init}\\u_{pred}\\\tilde{w}_{pred}\end{bmatrix}\;\label{eqn:rb_deepc_kkt},
\end{align}
\end{subequations}
where \change{$\kappa$ is the dual variable of~\eqref{eqn:rb_deepc_og_lower_cons} and}
\begin{subequations}
\begin{align}
    H&:=\begin{bmatrix}
    \Han_{L,init}(u_\textbf{d})\\\Han_{L,init}(\tilde{w}_\textbf{d})\\\Han_{L,pred}(u_\textbf{d})\\\Han_{L,pred}(\tilde{w}_\textbf{d})
    \end{bmatrix}\\
    M &:= \begin{bmatrix}
        \Han_{L,init}^\top(y_\textbf{d})\Han_{L,init}(y_\textbf{d})+\mathcal{E}_g &H^\top\\
        H&\textbf{O}
        \end{bmatrix}\;.\label{eqn:rb_deepc_kkt_mat}
\end{align}
\end{subequations}
\end{lemma}
\vspace{1.2em}
\begin{proof}
Note that the uncertain lower level problem in~\eqref{eqn:rb_deepc_og} is strongly convex and therefore can be equivalently represented by its KKT system~\cite[Chapter 4]{dempe2002foundations}. By replacing $\sigma_l$ by $\Han_{L,init}(y_\textbf{d})g_l-y_{init}$, the Lagrangian of the lower level problem is
\begin{align*}
    \mathcal{L}(g) = \frac{1}{2}\lVert \Han_{L,init}(y_\textbf{d})g&-y_{init}\rVert^2+\change{\frac{1}{2}}g^\top\mathcal{E}_g g\\
    &+\kappa^\top (Hg-\begin{bmatrix}
    u_{init}\\\tilde{w}_{init}\\u_{pred}\\\tilde{w}_{pred}
    \end{bmatrix})\;,
\end{align*}
where $\kappa$ is the dual variable of the equality constraint. Based on this, we have the stationary condition of the KKT system
\begin{align*}
    \change{\frac{\partial \mathcal{L}(g)}{\partial g}^\top} = (\Han_{L,init}(y_\textbf{d})^\top&\Han_{L,init}(\change{y_\textbf{d}})+\mathcal{E}_g)g +H^\top\kappa\\&-\Han_{L,init}(y_\textbf{d})^\top y_{init} = 0\;.
\end{align*}
By recalling the primal feasibility condition
\begin{align*}
    Hg = \begin{bmatrix}
    u_{init}\\\tilde{w}_{init}\\u_{pred}\\\tilde{w}_{pred}
    \end{bmatrix}\;,
\end{align*}
we get the uncertain KKT matrix $M$ in~\eqref{eqn:rb_deepc_kkt_mat}. \change{Finally, by Assumption~\ref{ass:PE}, $M$ is full-rank and hence invertible~\cite[Chapter 16]{nocedal2006numerical}. This leads to the uncertaint KKT equation~\eqref{eqn:rb_deepc_kkt},} which concludes the proof.
\end{proof}

On top of the single-level problem~\eqref{eqn:rb_deepc}, we further enforce causality on the decision variable $K$ through a lower-block triangular structure~\cite[Chapter 5.1]{lofberg2003minimax}:
\begin{align}\label{eqn:causal_fb}
    K = \begin{bmatrix}
    \textbf{O}&\textbf{O}&\textbf{O}&\dots&\vdots\\
    K_{2,1}&\textbf{O}&\textbf{O}&\dots&\vdots\\
    K_{3,1}&K_{3,2}&\textbf{O}&\dots&\vdots\\
    \vdots&\ddots&\ddots&\ddots&\vdots\\
    K_{n_h,1}&K_{n_h,2}&K_{n_h,3}&\dots&\textbf{O}
    \end{bmatrix}\;.
\end{align}
In particular, causality means that the $i$-th step of the future process noise can only change the events happening later than it, which only includes the $i+1$-th to the $n_h$-th components in $u_{pred}$.

\begin{remark}
When the feasible sets $\mathcal{\tilde{U},Y}$ are polytopic, the robust optimization problem~\eqref{eqn:rb_deepc} can be solved by a standard dualization procedure~\cite{ben2009robust}.
\end{remark}

\subsection{Discussion}\label{sect:compare}
\change{
We would wrap up this Section~\ref{sect:LTI_deepc} by a comparison between the proposed scheme and existing single-level schemes. Based on Willems' fundamental lemma~\ref{lem:funda}, a data-driven control scheme has been proposed in~\cite{coulson2019data,markovsky2007linear}
\begin{subequations}\label{eqn:deepc}
\begin{align}
        \min\limits_{\substack{y_{pred},u_{pred}\\g,\sigma}} \;&\; \underbrace{J(y_{pred},u_{pred})}_{(a)}+\underbrace{\eta_g \lVert g\rVert+\eta_\sigma \lVert \sigma \rVert}_{(b)}\label{eqn:deepc_obj}\\
        \text{s.t.} \;
        &\;\begin{bmatrix}
    \Han_{L,init}(y_\textbf{d})\\\Han_{L,init}(u_\textbf{d})\\\Han_{L,pred}(u_\textbf{d})\\\Han_{L,pred}(y_\textbf{d})
    \end{bmatrix}g=\begin{bmatrix}
        y_{init}+\sigma\\u_{init}\\u_{pred}\\y_{pred}\\
    \end{bmatrix}\label{eqn:deepc_pred}\\
        &\; u_{pred}\in \mathcal{U}\;,\;y_{pred}\in \mathcal{Y}\;,\nonumber
\end{align}
\end{subequations}
where $\eta_g$ and $\eta_\sigma$ are user-defined parameters and other components are similar to those defined in the proposed scheme~\eqref{eqn:rb_deepc_og}. For the sake of clarity, we neglect the process noise in this subsection. By comparing its objective~\eqref{eqn:deepc_obj} with the objective function in the lower level problem (i.e. Equation~\eqref{eqn:rb_deepc_og_lower}), one can see that the last two terms $(b)$ can be understood as the penalty on prediction error. These regularization terms are studied in~\cite{coulson2019regularized,dorfler2022bridging}. In particular, \cite{dorfler2022bridging} shows that the first term in $(b)$ is linked to the objective function used in the standard system identification procedure. Therefore, Problem~\eqref{eqn:deepc} can be understood as a bi-objective optimization problem, whose loss function tries to balance the prediction accuracy and control performance. Such a trade-off between these two objectives is modelled into the user-defined weights $\eta_g$ and $\eta_\sigma$. However, based on our experiments and the results reported in~\cite{elokda2021data,coulson2019data,dorf2019talk,huang2019decentralized}, the tunning of $\eta_g$ and $\eta_\sigma$ usually requires exhaustive search and is in general non-trivial. Instead of balancing the prediction accuracy and the control performance in a single objective function~\eqref{eqn:deepc_obj}, the proposed scheme couples them hierarchically in a bi-level optimization problem~\eqref{eqn:rb_deepc_og}. This follows an intuitive logic applied in predictive control: the control is decided based on an accurate output prediction. In the proposed scheme~\eqref{eqn:rb_deepc_og}, the prediction accuracy is optimized directly by the lower level problem. Thus, the prediction accuracy of the output trajectory is always guaranteed without sacrificing the control optimality (i.e. upper level objective). On the other hand, the prediction accuracy of the single-level scheme~\eqref{eqn:deepc} might be compromised due to the trade-off between control performance and the prediction accuracy (i.e. terms $(a)$ and $(b)$ in objective~\eqref{eqn:deepc_obj} respectively). To better see why the proposed scheme is preferable, one can consider a special case where $\mathcal{Y}=\mathbf{0},\;\mathcal{U}=\mathbf{0}$ and $u_{init}=\mathbf{0}$. In this case, problem~\eqref{eqn:deepc} has a non-empty solution set as $g=0$ is feasible regardless of the value of $y_{init}$. This solution corresponds to an output trajectory that can jump to the origin in one step with zero input regardless of the initial state. If we do not consider the case where $A_i=\mathbf{O}\;\forall\;i$ (see dynamics~\eqref{eqn:dyn_varying}), the underlying dynamics are not able to follow this predicted trajectory for arbitrary $y_{init}$, and hence the prediction can be inconsistent in the single-level scheme~\eqref{eqn:deepc}. In comparison, because the prediction accuracy is independently ensured by the lower level problem, the proposed scheme~\eqref{eqn:rb_deepc_og} will be infeasible unless the initial state is at the origin.} \changeTwo{In practice, the corner case given above barely happens, as reported by~\cite{dorf2019talk,coulson2019regularized2}, it is always possible to tune the $\eta_g$ and $\eta_\sigma$ to achieve a desirable closed loop performance. Note that, when the regularization term $(b)$ in~\eqref{eqn:deepc_obj} is not quadratic, its corresponding bi-level will be more difficult to solve numerically, as there is no convex single level reformulation. Hence, whether one uses the proposed bi-level structure or not depends on the specific application.}

\begin{remark}
\change{The idea of bi-level optimization is also presented in~\cite{dorfler2022bridging} and in subspace predictive control (SPC)~\cite{favoreel1999spc}. In these previous works, the lower level problem defines a system identification problem based on the historical data (i.e. $\{u_\mathbf{d},y_{\mathbf{d}}\}$). These approaches try to identify one single model, which is a preprocessed historical dataset in~\cite{dorfler2022bridging} (see e.g. Problem (5), (20) in~\cite{dorfler2022bridging}) and an ARX model in SPC. Thus, in this identification setup, the treatment of the noise presented in the online measurements (e.g. noise in $y_{init}$) is independent of the treatment of the measurement noise presented in the historical data (i.e. the identification problem). On the contrary, our proposed scheme deals with these two sources of measurement noise in a more unified way. Firstly, because of the noise in the historical data, the representation of the underlying dynamics is uncertain. In the proposed scheme, all the possible representations are considered (i.e. the uncertain Hankel matrices considered in Section~\ref{sect:wasser_bound}). Secondly, both the uncertain representations and the online measurement noise are treated in one single problem via a Wasserstein distance upper bound (Corollary~\ref{cor:wasser_bound}). Due to these differences, we intentionally call the lower level problem in the proposed scheme a ``trajectory prediction" problem but not an ``identification" problem.}
\end{remark}

\begin{remark}
\change{Different from our analysis, \cite{coulson2019regularized} studies a general setup that further includes unknown process noise. \cite{coulson2019regularized} applies the Wasserstein distance to show that the regularization on $g$ (first term of $(b)$ in~\eqref{eqn:deepc_obj}) is related to the minimization of the conditional value at risk (CVaR) from a distributionally robust viewpoint.}
\end{remark}

\begin{remark}
\change{The objective function $J(y_{pred},u_{pred})$ has different choices, such as the robust objective $J(y_{pred},u_{pred})=\max\limits_{\tilde{w}_{pred}} \tilde{J}(y_{pred},u_{pred})$. However, unless $\tilde{J}(\cdot,\cdot)$ is linear, this robust objective is non-convex, and therefore does not meet our requirement (Section~\ref{sect:stage}). Notice that the certainty equivalence is widely adopted in building applications, it is therefore reasonable to optimize the nominal performance (i.e. $J(y_{pred},u_{pred})=\tilde{J}(\overline{y}_{pred},\overline{u}_{pred})$).}
\end{remark}

\section{Heuristic Time-varying Extension}\label{sect:LTV_deepc} 
In this section, we will adapt the controller~\eqref{eqn:rb_deepc_og} to LTV dynamics~\eqref{eqn:dyn_varying} by two heuristics.  The modifications are summarized as follows:
\begin{enumerate}
    \item \textbf{Modification of the lower level objective:} $\mathcal{E}_g$ is a diagonal matrix $\text{diag}(\eta_{g,1},\dots,\eta_{g,n_c})$ and the weight sequence $\{\eta_{g,i}\}_{i=1}^{n_c}$ is decreasing.  Recall that the $i$-th entry of $g$, $g_i$, is the weight of the $i$-th column in $H$ and $\Han_{L,init}(y_\textbf{d})$ used for prediction. Thus, a non-uniform penalty on $g$ can be used to model our preference  of using recent data for trajectory prediction, and the decreasing diagonal elements in $\mathcal{E}_g$ reflect this preference.
    \item \textbf{Adaptation of the measured dataset: }The data are updated online to capture the latest dynamics from the system. In particular, Hankel matrices are updated by appending new input/output measurements on the right side of the Hankel matrices and by discarding old data on the left side.
\end{enumerate}
\change{Note that the second heuristic is tailored for slowly time-varying systems. This is the case for building applications, whose variations are usually seasonal. This section will discuss how to update the dataset in a way that can robustly guarantee the data quality (Section~\ref{sect:act}), and that has a scalable update computation (Section~\ref{sect:numerical}).}
\change{It is worth mentioning that updating dataset online is also used in~\cite{berberich2022linear} to learn the linearized model around different operating points online.}

\subsection{Active Excitation}\label{sect:act}
In this part, we will discuss the selection of the $\tilde{\mathcal{U}}$ and $\tilde{\mathcal{W}}$ in~\eqref{eqn:rb_deepc_og}. Recall that persistent excitation is the key assumption required for Willems' fundamental lemma to apply. \change{As the persistent excitation condition is not explicitly considered in the predictive control problem~\eqref{eqn:rb_deepc_og}, thus the control input excitation may become impersistent in closed loop. In this case, updating Hankel matrices with these latest I/O sequence is not reasonable.} For example, in building control, if the outdoor temperature and/or solar irradiation are near the building's equilibrium point, no extra heating/cooling is needed when the energy consumption is aimed to be minimized. In this case, \change{the long-term zero-valued control input will not excite the system persistently.} To accommodate this issue, we introduce a robust active excitation scheme, which perturbs the control input applied at time $i$ by a random excitation signal
\begin{align}\label{eqn:u_act}
    u_i = \underbrace{\overline{u}_{pred,1|i}}_{(a)}+\underbrace{{u}_{e,i}}_{(b)}\;,
\end{align} 
where $\overline{u}_{pred,1|i}$ is the first element of $\overline{u}_{pred}$ determined by a predictive control \change{problem~\eqref{eqn:rb_deepc_og}} solved at time $i$. In this decomposition, the term $(a)$ is determined by the predictive control \change{problem~\eqref{eqn:rb_deepc_og}} with some specific choice of $\tilde{\mathcal{U}}$ and $\tilde{\mathcal{W}}$ (see Algorithm~\eqref{alg:active_deepc} below), and the term $(b)$ is a bounded excitation input, which is unknown to the decision process of $\overline{u}_{pred,1|i}$. More specifically, from the viewpoint of $u_{pred}$, $u_e$ is an uncontrolled, but measurable, process noise and the underlying time-varying linear dynamics therefore becomes
\begin{align*}
    x_{i+1} &= A_ix_i+B_iu_{pred,1|i}+\begin{bmatrix}E_i &B_i\end{bmatrix}\begin{bmatrix}w_i\\u_{e,i} \end{bmatrix}\;,
\end{align*}
where the excitation input is randomly sampled from a user-defined compact set as $u_e\in\mathcal{U}_e\subset \mathcal{U}$. The process noise $\tilde{w}$ is accordingly augmented to $\begin{bmatrix}w_{pred}^\top&u_e^\top\end{bmatrix}^\top$, and the uncertainty set $\tilde{\mathcal{W}}$ in~\eqref{eqn:rb_deepc_og_wcons} is augmented to $\left(\overline{w}\oplus\mathcal{W}\right)\times \mathcal{U}_e$. As a result, $\tilde{w}_{init}$ is set to $\begin{bmatrix}w_{init}^\top&\mathbf{0}^\top\end{bmatrix}^\top$. Meanwhile, due to an extra excitation signal in~\eqref{eqn:u_act}, the feasible set of the control input $\tilde{\mathcal{U}}$ is tightened to $\mathcal{U}\ominus\mathcal{U}_e$.

In practice, this active excitation mechanism sacrifices the flexibility of the control input for data quality. If not necessary, we should set $\tilde{\mathcal{U}}=\mathcal{U}$ and $\tilde{\mathcal{W}}=\mathcal{W}$ to generate control inputs and deactivate the active excitation scheme. We summarize the general algorithm of the proposed controller in Algorithm~\ref{alg:active_deepc}, which automatically selects the uncertainty set $\tilde{\mathcal{W}}$ and input feasible set $\tilde{\mathcal{U}}$.
\begin{algorithm}
\caption{$\enspace$}\label{alg:active_deepc}
\begin{algorithmic}
\WHILE{true}
    \STATE Measure $y_i,w_i$ and update Hankel matrices.
    \STATE $\tilde{\mathcal{U}}\gets\mathcal{U},\tilde{\mathcal{W}}\gets \overline{w}\oplus\mathcal{W}$
    \STATE Solve \change{problem~\eqref{eqn:rb_deepc_og}} to get $\overline{u}_{pred|i}$
    \IF{\change{input excitation $\{u_\textbf{d},\overline{u}_{pred|i}\}$ \changeTwo{is not persistently exciting}}}
        \STATE $\tilde{\mathcal{U}}\gets\mathcal{U}\ominus \mathcal{U}_e,\;\tilde{\mathcal{W}}\gets \left(\overline{w}\oplus\mathcal{W}\right)\times \mathcal{U}_e$
        \STATE Solve \change{problem~\eqref{eqn:rb_deepc_og}} to get $\overline{u}_{pred|i}$
        \STATE Sample $u_{e,i}$ from $\mathcal{U}_e$
        \STATE Apply $u_i=\overline{u}_{pred,1|i}+u_{e,i}$
    \ELSE
        \STATE Apply $u_i=\overline{u}_{pred,1|i}$
    \ENDIF 
    \STATE $i\gets i+1$
\ENDWHILE
\end{algorithmic}
\end{algorithm}

In this algorithm, the problem~\eqref{eqn:rb_deepc} is first solved without active excitation. \change{If its nominal solution $\overline{u}_{pred}$ is not expected to be persistently excited,} the problem~\eqref{eqn:rb_deepc} is re-solved considering the active excitation scheme. Note that the computational cost of checking the rank condition is $O(n_c^3)$~\cite{stewart1998matrix}, and may not be affordable for online computation. This can be replaced by some effective heuristics. In building control, \change{the control input excitation becomes impersistent mainly} when the control input is zero \change{(e.g. when the cooling/heating is turned off)}, and thus the persistence excitation condition can be heuristically replaced by checking whether the nominal predictive input is \change{near to a zero valued sequence up to some user-defined tolerance}. \change{This heuristic is \changeTwo{particularly} useful in building control. Because when input is not zero-valued, the stochastic property of the process noise (e.g. solar radiation and outdoor weather) will cause random fluctuation in the closed-loop input trajectory, and the persistent excitation condition is in turn satisfied. On top of this aspect, due to the stochastic property of the process noise, the process noise is usually persistently excited.}

\begin{remark}
In general, generating a persistently excited control input while considering control performance is challenging, as the persistent excitation condition depends on the rank of $\Han_L(u_\textbf{d})$, which turns the optimization problem into a challenging non-smooth non-convex optimization~\cite{marafioti2014persistently}. It is noteworthy that~\cite{faradonbeh2020input} also perturbs the nominal control input to guarantee persistent excitation, however, their result has no guarantee of robust constraint satisfaction.
\end{remark}
\begin{remark}
Due to the causality constraints in~\eqref{eqn:causal_fb}, the matrix $K$ in~\eqref{eqn:rb_deepc_og_ucons} cannot instantaneously counteract the excitation signal $u_e$ with a $K=\begin{bmatrix}K_w&-I\end{bmatrix}$, which is non-causal.
\end{remark}

\subsection{Numerical Details}\label{sect:numerical}
In our proposed adaptive robust controller, the Hankel matrices are updated online with the real-time measurement of $u_i,y_i,w_i$ (Section~\ref{sect:LTV_deepc} and Algorithm~\ref{alg:active_deepc}). Meanwhile, a numerically efficient reformulation of the robust problem~\eqref{eqn:rb_deepc} requires an explicit evaluation of matrix inversion $M^{-1}$ in~\eqref{eqn:rb_deepc_kkt} at each update. More specifically, when the feasible sets $\mathcal{\tilde{U},Y}$ and the uncertainty set $\mathcal{\tilde{W}}$ are polytopic or ellipsoidal, the dualization/explicit upper bound of the robust inequality constraint depends on the matrix inversion $M^{-1}$. However, the computational cost of $M^{-1}$ is $O((n_c+n_r)^3)$~\cite{stewart1998matrix} \change{with $n_r$ and $n_c$ the number of rows and columns in the matrix $H$}, which is roughly cubic in the size of the data set. Thus, direct inverse of $M$ online is not scalable. \change{This is particularly important for building applications, because the computing unit in building applications is usually of lower performance~\cite{sturzenegger2014model}}. We therefore propose to apply two linear algebraic techniques to resolve this computational bottleneck. 

Notice that the dual variable $\kappa$ in the lower problem does not affect the upper level problem. For the sake of compactness, we denote $M_{1,1}:=\Han_{L,init}^\top(y_\textbf{d})\Han_{L,init}(y_\textbf{d})+\mathcal{E}_g$ \change{and $M_{\mathrm{sch}}:=(HM_{1,1}^{-1}H^\top)^{-1}HM_{1,1}^{-1}$.} By matrix inversion of a block-structured matrix, we have
\change{
\begin{align*}
    M^{-1} = \begin{bmatrix}
    M_{1,1}^{-1}-M_{1,1}^{-1}H^\top M_{\mathrm{sch}}& \changeTwo{M_{\mathrm{sch}}^{\top}}\\
    \changeTwo{M_{\mathrm{sch}}}& -(HM_{1,1}^{-1}H^\top)^{-1}
    \end{bmatrix}\;.
\end{align*}}

We can therefore replace the constraint~\eqref{eqn:rb_deepc_kkt} by
\change{\begin{align*}
    g(\tilde{w}_{pred}) = M_{\mathrm{top}} \begin{bmatrix}\Han_{L,init}(y_\textbf{d})^\top y_{init}\\u_{init}\\\tilde{w}_{init}\\u_{pred}\\\tilde{w}_{pred}\end{bmatrix}\;,
\end{align*}}
where 
\begin{align*}
     M_{\mathrm{top}}:=\begin{bmatrix}
      M_{1,1}^{-1}-M_{1,1}^{-1}H^\top M_{\mathrm{sch}}& M_{\mathrm{sch}}
    \end{bmatrix}\;.
\end{align*}

With the aforementioned modification, the computational cost is lowered to $O(n_c^3)$, whose computational bottleneck lies at the inversion of $M_{1,1}$. We further lower the computational cost by the Woodbury matrix identity,
\change{\begin{align}\label{eqn:inv_woodbury}
\begin{split}
    &M_{1,1}^{-1} \\
    &= (\Han_{L,init}^\top(y_\textbf{d})\Han_{L,init}(y_\textbf{d})+\mathcal{E}_g)^{-1}\\
    &= \mathcal{E}_g^{-1}-\mathcal{E}_g^{-1}\Han_{L,init}^\top(y_\textbf{d}) {M_{\mathrm{mid}}}^{-1}\Han_{L,init}(y_\textbf{d})\mathcal{E}_g^{-1}\;,
\end{split}
\end{align}
where $M_{\mathrm{mid}}:=I_{m}+\Han_{L,init}(y_\textbf{d})\mathcal{E}_g^{-1}\Han_{L,init}^\top(y_\textbf{d})$.} As $\mathcal{E}_g$ is diagonal with a simple and explicit inversion, the major computation cost happens at the inversion of a size $m$ matrix $M_{mid}$, where $m:= t_{init} n_y $. Thus, the computational cost of $M_{1,1}^{-1}$ and $M_{\mathrm{top}}$ is lowered to $O(m^3)$, which is fixed and independent of the number of columns in the Hankel matrices (\textit{i.e.} roughly the size of the data-set). 
\begin{remark}
Note that the KKT matrix is usually ill-conditioned~\cite[Chapter 16]{nocedal2006numerical}. Replacing the full matrix inversion in~\eqref{eqn:rb_deepc_kkt} with the proposed techniques can improve the numerical stability. Because only the matrix inversion of $\mathcal{E}_g$ and $M_{\mathrm{mid}}$ in~\eqref{eqn:inv_woodbury} are evaluated, and these two matrices are well-conditioned.
\end{remark}

\section{Numerical Results}\label{sect:simulation}
In this part, we will demonstrate that our proposed scheme shows comparable performance against some model-based methods in both LTI and LTV systems. 

\subsection{Multi-zone Building Model}\label{sect:2d_demo}
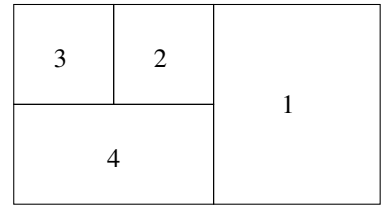
\begin{figure}[!htb]
    \centering
    \begin{tikzpicture}
    \draw [draw=black] (0,0) rectangle (.3\linewidth,.15\linewidth);
    \draw [draw=black] (0,.15\linewidth) rectangle (.15\linewidth,.3\linewidth);
    \draw [draw=black] (.15\linewidth,.15\linewidth) rectangle (.3\linewidth,.3\linewidth);
    \draw [draw=black] (.3\linewidth,0) rectangle (.55\linewidth,.3\linewidth);
    \node at (0.15\linewidth,.07\linewidth) {4};
    \node at (0.07\linewidth,.22\linewidth) {3};
    \node at (0.22\linewidth,.22\linewidth) {2};
    \node at (0.412\linewidth,.15\linewidth) {1};
    \end{tikzpicture}
    \caption{Schematic diagram of the multi-zone building}
    \label{fig:multizone_scheme}
\end{figure}

\change{It should be noted that we are not claiming that the data-driven approach outperforms all model-based approaches, as it is definitely possible to tune a better model based method, such as considering the uncertainty of the identified parameters or using a more complex estimator/controller. We only aim to show that the proposed approach has comparable performance against a model-based method, but without requiring a model. We therefore select a standard model-based controller design pipeline that we believe is reasonable, and we compare this standard scheme against our proposed scheme in this example. We considered an LTI multi-zone building model reported in~\cite{belic2021detailed} (index of the rooms are shown in schematic diagram Figure.~\ref{fig:multizone_scheme}. Due to the space limit, the parameters of the model (i.e. $A,\;B,\;C,\;D,\;E$ matrices) are included in the supplementary material on Github\footnote{\href{https://github.com/YingZhaoleo/Building_results}{https://github.com/YingZhaoleo/Building\_results}}. In this multi-zone building (Figure~\ref{fig:multizone_scheme}), room 4 is the corridor linking a large warehouse (room 1) and two offices (room 2 and 3). The indoor temperature of room 1 is controlled by an independent HVAC system, while another HVAC controls the temperature of all other rooms (i.e. $u\in\R^2$). Only the indoor temperature of these four rooms are measured (i.e. $y\in\R^4$), while the underlying model is 13 dimensional including the wall temperature (i.e. $x\in\R^{13}$). Process noise are outdoor temperature and solar radiation (i.e. $w\in\R^2$), and real weather data is used for the closed-loop simulation. The sampling time of this discrete time model is 15 minutes. Recall the system dynamics~\eqref{eqn:dyn_varying}, two different level of unknown measurement noise $v$ are considered to show the reliability of the proposed scheme. We remind the reader that only the measurement $y_i$ is available to both schemes and $\overline{y}_i$ is unknown to both schemes (See the dynamics~\eqref{eqn:dyn_varying}).}

\change{
In the ``standard" scheme, the model is identified by a subspace identification algorithm and the state estimation is done by a Kalman filter\footnote{We use the commands \textsc{n4sid} and \textsc{kalman} in \textsc{Matlab} to do subspace identification and Kalman filter design respectively}. Following this, a robust model predictive controller with linear feedback~\cite{lofberg2003minimax} is used in the ``standard" scheme to generate control inputs\footnote{We used the code template of ``Approximate closed-loop minimax solution'' in \href{url}{https://yalmip.github.io/example/robustmpc}.}
\begin{subequations}\label{eqn:rb_mpc}
\begin{align}
\begin{split}
\min\limits_{\substack{u_{pred},K_w\\y_{pred}}}& J(\overline{y}_{pred},\overline{u}_{pred})\\
    \text{s.t.}\;& x_{pred,0} = x_0\\
    &\forall\; i = 0,1\dots,n_h\\
    & x_{pred,i+1} = A_{id}x_{pred,i}+B_{id}u_{pred,i}\\
    &\qquad\qquad\qquad+E_{id}w_{pred,i}\\
    & y_{pred,i} = C_{id}x_{pred,i}+D_{id}u_{pred,i}\;\\
    & \forall\; w_{pred}\in\mathcal{W}\;,\; u_{pred}=\overline{u}_{pred}+K_ww_{pred}\\
    & u_{pred}\in\mathcal{U}\;,\;y_{pred}\in\mathcal{Y}\;,
\end{split}
\end{align}
\end{subequations}
where the parameters of $A_{id}\;,\;B_{id}\;,\;E_{id}\;,\;C_{id}\;,\;D_{id}$ come from system identification and $x_0$ is estimated with a Kalman filter. Note that the feedback control law $K_w$ is optimized by~\eqref{eqn:rb_mpc}. Other components, such as $J(\cdot,\cdot)$, $n_h$ and $\mathcal{U,Y,W}$, are identical to those used in the proposed robust data-driven scheme. In particular, the heating power (control input) $ 0 \;kW\leq u_{pred}\leq 4.5\;kW$, the indoor temperature for all the rooms are bounded within $20 ^\circ C\leq y_{pred}\leq 26 ^\circ C$ and the uncertainty of the solar radiation, outdoor temperature prediction is modelled by a tube around their nominal prediction. The radius of the tube are respectively $2 \;kW/m^2$ and $2 ^\circ C$, and the nominal prediction $\overline{w}_{pred,i}$ comes from the weather forecast. An indoor temperature control problem is considered:
\begin{align*}
    J(y_{pred},u_{pred}) = \sum\limits_{i=0}^{n_h} 10(\overline{y}_{pred,i+1}-y_{ref})^2+0.1\overline{u}_{pred,i}^2\;,
\end{align*}
where $y_{ref}$ is the set point of the indoor temperature and the prediction horizon is set to $n_h=8$, and the sampling time of this model $15$ minutes. Additionally, we set $t_{init}=5$ in the proposed robust scheme. To ensure a fair comparison, the same dataset is used for system identification, and for defining the Hankel matrices in the LTI case as well as the initial Hankel matrices in the LTV case. Four days of historical data is used, which includes $384$ data points. }

\input{data/multizone}
\begin{table}[h!]
    \centering
    \setlength{\tabcolsep}{3.2pt}
\begin{tabular}{c |c c c c c}
         \multicolumn{2}{c}{Room index} & 1  & 2 & 3 &4\\\hline
     \multirow{3}{*}{\shortstack{Constraint\\violation}}&\shortstack{Proposed\\ scheme} & \shortstack{0.67\%\\(3.12\%)} & \shortstack{0\%\\(0.14\%)} & \shortstack{0.40\%\\(2.76\%)} & \shortstack{0\%\\(0.078\%)}\\\cline{2-6}
     &\shortstack{Standard\\scheme} & \shortstack{2.78\%\\(11.46\%)} & \shortstack{0\%\\(0.99\%)} & \shortstack{3.28\%\\(10.27\%)} & \shortstack{0.42\%\\(3.77\%)}\\\hline
     \multicolumn{2}{c}{\shortstack{Averaged fitting\\ accuracy}}& \shortstack{93.17\%\\(77.60\%)}& \shortstack{93.07\%\\(73.43\%)} & \shortstack{90.72\%\\(71.32\%)} & \shortstack{92.58\%\\(83.93\%)}
\end{tabular}
    \caption{\change{Statistics of the constraint violation and the fitting accuracy of the identified models. The bracket number in each entry corresponds to the tests whose standard deviation of measurment noise is $0.3 ^\circ C$, the unbracketed ones are of $0.05 ^\circ C$ measurement noise standard deviation.}} 
    \label{tab:vio}
\end{table}

\change{
Two experiments with different levels of measurement noise are considered, whose standard deviation are respectively $0.05 ^\circ C$ and $0.3^\circ C$. These two cases correspond to a measurement error roughly bounded by $0.15 ^\circ C$ and $1^\circ C$ respectively. The results are shown in Figure~\ref{fig:multizone}}\footnote{To be fair, both controllers are subject to the same measurement noise and process noise during their online operation.} \change{and Table~\ref{tab:vio}. Each experiment carries out 50 Monte-Carlo runs. In each Monte-Carlo run, a new dataset is generated for system identification and for the definition of the Hankel matrices. The averaged fitting accuracy\footnote{\change{The fitting accuracy is generated by the \textsc{compare} command in \textsc{Matlab}.}} are summarized in Table~\ref{tab:vio} as well, which shows good modelling accuracy in the ``standard" scheme. Additionally, the constraint violation is calculated by
\begingroup\makeatletter\def\f@size{9.5}\check@mathfonts
\begin{align*}
     \frac{\text{number of steps where a constraint violation occurs}}{(\text{number of simulation steps}\times \text{number of Monte-Carlo runs})}\;.
\end{align*}
\endgroup}

\change{From the rows (a) and (b) in Figure~\ref{fig:multizone}, we can observe that the proposed scheme shows comparable performance against the model based scheme. Meanwhile, when the measurement noise is larger, the variance of the closed loop trajectory is higher in the data-driven scheme and the model based scheme. This also leads to more frequent constraint violations. While both controllers violate the constraints more frequently in this case, our proposed scheme offers a better constraint satisfaction guarantee. The constraint violation statistics are summarized in Table~\ref{tab:vio}, and we observe that our proposed controller offers much better constraint satisfaction. In the proposed scheme, constraint violations are caused by the reason discussed in Remark~\ref{rmk:pred_uncertain}. On the other hand, we believe that the constraint violation in the model-based method can be resolved by more complex methods such as constraint tightening, but there is currently no systematic way to achieve this. Last but not least, the in comparison with the single-level scheme, the proposed scheme shows lower variance in the closed-loop operation, which leads to a higher constraint satisfaction when the measurement noise is higher.}

\change{
In the second test, we consider that the building dynamics varies slowly on a weekly basis:
\begin{align*}
    &A_i = A+0.02\sin\left(\frac{i\pi}{336}\right)I_{13}\\ 
    &B_i = B,\; C_i = C,\; D_i = D,\; E_i = E,\;\forall\;i\;.
\end{align*}
Our proposed scheme is compared against a model based adaptive method, where a recursive least square (RLS) estimator updates the parameter of an ARX model:
\begin{align*}
    y_i = \sum\limits_{j=1}^{t_{init}} \theta_{y,j}^\top y_{i-j}+\theta_{u,j}^\top u_{i-j}+\theta_{w,j}^\top w_{i-j}\;,
\end{align*}
where $\theta_{y,j}\in\R^4,\;\theta_{u,j}\in\R^2,\;\theta_{w,j}\in\R^2$.
The model estimated by RLS is used in the following robust MPC problem:
\begin{align}\label{eqn:rls_rb_mpc}
\begin{split}
    \min\limits_{\substack{u_{pred},K_w\\y_{pred}}} &\max_{w_{pred}}J(y_{pred},u_{pred})\\
    \text{s.t.}\;& x_{pred,0} = x_0\\
    &\forall\; i = 0,1\dots,n_h\\
    &y_{pred,i} = \sum\limits_{j=1}^{t_{init}} \theta_{y,j}y_{pred,i-j}+\theta_{u,j}u_{pred,i-j}\\
    &\qquad\qquad+\theta_{w,j}w_{pred,i-j}\\
    & \forall\; w_{pred}\in\mathcal{W}\;,\; u_{pred}=\overline{u}_{pred}+K_ww_{pred}\\
    & u_{pred}\in\mathcal{U}\;,\;y_{pred}\in\mathcal{Y}\;,
\end{split}
\end{align}
where the parameters $\theta_y,\theta_u$ and $\theta_w$ are updated by the RLS estimator. Other settings are the same as the previous experiments. In particular, the RLS estimator has a forgetting factor of $0.98$ and it is initialized by the data used to build the initial Hankel matrices of the proposed scheme. Two experiments with different levels of measurement noise standard deviation are conducted. The results are shown in rows (c) and (d) in Figure~\ref{fig:multizone} and the constraint violation statistics are given in Table~\ref{tab:vio_varying}. From rows (c) and (d) in Figure~\ref{fig:multizone}, we can see that both approaches perform the tracking task properly, and the proposed scheme shows comparable performance against the RLS-MPC approach. One major observation is that the proposed scheme shows better constraint satisfaction against the RLS-MPC method. Indeed, it is possible to consider the uncertainty generated by the RLS estimator to improve the robustness of the model-based approach, however, it turns out to be a non-convex robust optimization problem and there is no standard approach to solve this problem.}

\begin{table}[h!]
    \centering
    \setlength{\tabcolsep}{3.2pt}
\begin{tabular}{c |c c c c c}
         \multicolumn{2}{c}{room index} & 1  & 2 & 3 &4\\\hline
     \multirow{2}{*}{\shortstack{Constraint\\violation}}&\shortstack{Proposed\\ scheme} & \shortstack{1.64\%\\(6.99\%)} & \shortstack{0\%\\(0.49\%)} & \shortstack{2.70\%\\(7.65\%)} & \shortstack{0.47\%\\(2.25\%)}\\\cline{2-6}
     &\shortstack{RLS-MPC\\scheme} & \shortstack{8.23\%\\(22.54\%)} & \shortstack{0.37\%\\(3.08\%)} & \shortstack{10.16\%\\(22.21\%)} & \shortstack{0.69\%\\(11.11\%)}
\end{tabular}
    \caption{\change{Statistic of the constriant violation. The bracket number in each entry corresponds to the tests whose standard deviation of measurment noise is $0.3 ^\circ C$, the unbracketed ones are of $0.05 ^\circ C$ measurement noise standard deviation.}} 
    \label{tab:vio_varying}
\end{table}

\section{Experiment}\label{sect:experiment}
\textcolor{black}{This part presents our real-world experiments conducted at a conference building on the EPFL campus.}

\subsection{Experimental Setup}\label{sect:exp_setup}

\begin{figure}[h]
    \centering
    \includegraphics[width=0.8\linewidth]{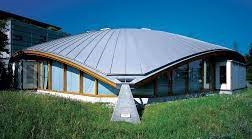}
    \caption{The Polydome}
    \label{fig:polydome_ext}
\end{figure}

\textcolor{black}{The building control experiment is conducted on an entire building, which is a freestanding $600 m^2$ single-zone building on the EPFL campus, called the \textit{Polydome}. It is regularly used for lectures/exams and accommodates up to 200 people (Figure~\ref{fig:polydome_ext}).}

In the presented experiments, the indoor temperature measurement is the noisy output measurement $y$, the active electrical \change{energy} consumption of the HVAC system is the input $u$,  and the weather conditions (outdoor temperature and solar radiation) are the process noise $ w = \begin{bmatrix}w_1&w_2\end{bmatrix}$. The sampling period $T_s$ is 15 minutes and the structure of the control system is depicted in Figure~\ref{fig:ctrl_sys}, where the arrows indicate the direction of data transmission. The system consists of five main components
\begin{itemize}
    \item \textbf{Sensors}: Four Z-wave \textsc{Fibaro Door/Window Sensor v2} are put in different locations in the \textit{Polydome} to measure the indoor temperature (path $(a)$). Every five minutes, the temperature measurements are sent to the database through a wireless Z-wave network (path $(c)$). The average value of the four measurements is used as the indoor temperature.  The active power consumption of the HVAC system is measured via an EMU 3-phase power meter~\cite{emu} (path $(b)$).
    \item \textbf{Database}: We use \textsc{InfluxDB}~1.3.7~\cite{influxdb} to log the time-series data, which records the measurements from sensors (path $(c)$), the control input computed from the PC (path $(d)$) and the historical weather (path $(f)$).
    \item \textbf{Weather API}: We use \textsc{Tomorrow.io}~\cite{tomorrow}, which provides both historical and current measurements of solar radiation and outdoor temperature to the database (path $(f)$). It also provides the forecast of solar radiation and outdoor temperature to the PC to solve the predictive control problem (path $(g)$).
    \item \textbf{Controller}: The controller is implemented in \textsc{MATLAB}, interfacing \textsc{YALMIP}~\cite{lofberg2004yalmip}, which fetches historical data from the database to build/update the Hankel matrices (path $(d)$), and acquires weather forecasts from the weather API (path $(g)$). It runs Algorithm~\ref{alg:active_deepc} to generate the control input. This control signal is transmitted to the HVAC system via the serial communication protocol, Modbus~\cite{swales1999open}.
    \item \textbf{HVAC}: A roof-top HVAC unit (series No: AERMEC RTY-04 heat pump) is used for heating, cooling and ventilation. Its heating and cooling units are different, consisting of two compressors for heating and one compressor for cooling respectively.
\end{itemize}
\tikzstyle{block} = [rectangle, draw, fill=blue!20, 
    text width=5em, text centered, rounded corners, minimum height=3em]

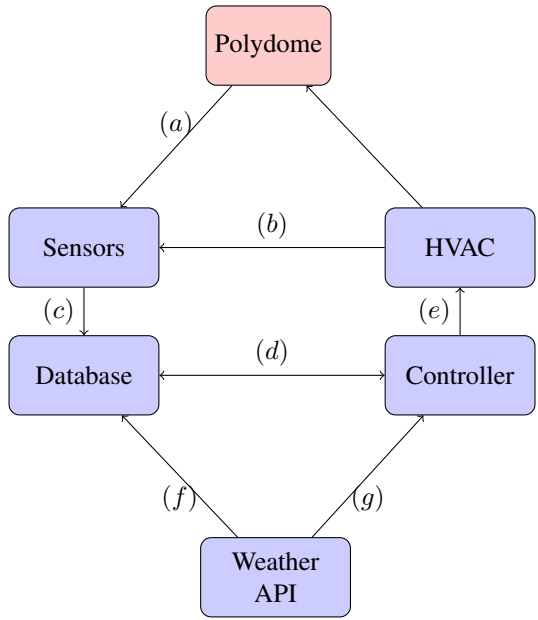
\begin{figure}[h]
    \centering
    \begin{tikzpicture}
    [node distance = 1.7cm, auto]
    \node [block] (sensor) {Sensors};
    \node [block, below of=sensor] (db) {Database};
    \node [block, right of=sensor,node distance=5cm] (HVAC) {HVAC};
    \node [block, below of=HVAC] (PC) {Controller};
    \node [rectangle,draw,fill=red!20,text centered, rounded corners, minimum height=4em,above left of = HVAC,minimum height=3em,node distance = 3.8cm,xshift = .14cm](Polydome){Polydome};
    \node[block, below right of = db,node distance = 3.8cm,xshift = -.14cm](weather){Weather API};
    \draw[thin,->] (sensor)--node [midway, left,text centered] {$(c)$}(db);
    \path[draw,<->] (db)--node [midway,above,text centered]{$(d)$}(PC);
    \draw[thin,->] (PC)--node [midway, left,text centered] {$(e)$}(HVAC);
    \path[draw,->](HVAC)--node [midway, above,text centered] {$(b)$}(sensor);
    \draw[thin,->](HVAC)--(Polydome);
    \path[draw,->](Polydome)--node[midway,above,text centered]{$(a)$}(sensor);
    \path[draw,->](weather)--node[midway,below,text centered]{$(f)$}(db);
    \path[draw,->](weather)--node[midway,below,text centered]{$(g)$}(PC);

    \end{tikzpicture}
    \caption{Structure of the building control system}
    \label{fig:ctrl_sys}
\end{figure}

The HVAC system is shipped with an internal hierarchical controller, which includes:
\begin{itemize}
    \item \textbf{Mode scheduler:} The scheduler determines whether the HVAC is in heating or cooling mode, and we are \textbf{not} authorized to access this scheduler.
    \item \textbf{Temperature controller:} The indoor temperature is controlled by a bang-bang controller that compares the set-point temperature and the return-air (indoor) temperature with a dead-band of $1 \degree C$. For example, in the heating mode, if the return-air temperature is $1 \degree C$ lower than the set-point temperature, the heating will turn on and run at full power of $8.4\, kW$ until the set-point is reached. Vice versa for cooling, except that the electric power is $7\, kW$.
\end{itemize}

To map this controller to our proposed controller, we applied the following strategies:
\begin{itemize}
    \item As the cooling and heating modes of the HVAC system show different responses, two different I/O datasets for different modes are maintained/updated independently. The controller monitors the mode of the HVAC system and deploys the corresponding I/O dataset to build the proposed controller.
    \item Recall that the input used in our proposed scheme is electrical \change{energy} consumption. To achieve a desired power consumption, we convert this desired consumption to a set-point sequence: For example, if the HVAC is in heating mode, and a non-zero desired power consumption is planned, the controller will turn on the heating by giving a set-point that is $2\degree C$ above the return-air temperature, until the desired energy ($P_{el}T_{s}$) is reached (path $(b)$ in Figure~\ref{fig:ctrl_sys}). Then, the heating is turned off by setting the set-point to the return-air temperature. 
\end{itemize}

In our bi-level  predictive  control scheme, the Hankel matrices for both heating and cooling modes are built from 200 data-points, with $t_{init}=10$ and a prediction horizon $n_h=10$. The controller minimizes electrical power consumption with the following loss function
\begin{align*}
&J(y_{pred},u_{pred})=\sum_{i}^{N_{pred}} |\overline{u}_{i}| 
\end{align*}

To better distinguish the heating and cooling modes in our plots, we use a positive input value for the heating mode and a negative input for the cooling mode. We further enforce the following input constraint to model the maximal \change{energy} consumption of the heating/cooling.
\begin{align*}
    \left\{
     \begin{array}{rl}
	 0 \,kWh \; \leq u_i \leq \;1.5\,kWh, &		\text{heating mode} \\
	 -1.15 \, kWh\;\leq u_i \leq \;0 \, kWh, &		\text{cooling mode}\\
	\end{array}
	\right.
\end{align*}

Note that the HVAC unit consumes a constant $2.4 \, kWh$ of \change{energy} for ventilation, even without heating or cooling. The aforementioned input constraint excludes this basic ventilation power. The parameter $\mathcal{E}_g$ in~\eqref{eqn:rb_deepc} is set by \textsc{Matlab} command \textsc{diag(linspace(0.2,0.02,$n_c$))}.
The uncertainty set of the weather forecasts is estimated from an analysis of historical data as
\begin{align}\label{eqn:poly_tube}
    \mathcal{W}:=\left\{w_i\middle|
     \begin{bmatrix}
	 -1 \degree C \\
	 -50 W/m^2
     \end{bmatrix} \leq w_i \leq
     \begin{bmatrix}
	 1 \degree C \\
	 50 W/m^2
     \end{bmatrix}\right\}
\end{align}

\subsection{Experimental Results}
In this section, we describe four experiments that were conducted from May 2021 to June 2021. In particular, the first experiment shows the necessity of robust control and the second experiment shows the adaptivity to mode switching. The \change{third} one runs a 20-day experiment to show the adaptivity and reliability of the proposed scheme and the \change{fourth} one runs a 4-day experiment to compare the proposed scheme with the default controller. Meanwhile, recall that we use a negative control input to represent cooling and a positive control input for heating. Accordingly, and show the control input and system output (indoor temperature) within the same plot to better show the response from input to output.

\subsubsection{Experiment 1}

The first experiment includes two parts: a non-robust version of the proposed scheme (\textit{i.e. }$\mathcal{W}=\{\textbf{0}\},\;K=\textbf{O}$) and a robust version with the uncertainty tube given in~\eqref{eqn:poly_tube}. In this test, we consider a time-varying indoor temperature constraint with respect to office hours, which is relaxed during the night and is tightened to ensure occupant comfort during office hours.
\begin{align*}
    \left\{ \begin{array}{ll}
	21\degree C\le y_i\le 26\degree C,&		\text{from 8 a.m. to 6 p.m.} \\
	19\degree C\le y_i\le 30\degree C,&		\text{otherwise}\\
\end{array} \right.
\end{align*}

The non-robust experiment was conducted on $14^{th}$ May 2021 and the result is plotted in Figure~\ref{fig:exp_polydome_2}. The HVAC system was in heating mode throughout this experiment, and we can observe that \change{the input was 0 until the indoor temperature hit the lower bound at around 4 A.M..} Later, it started pre-heating the room to satisfy the office hours temperature constraint at around 6 A.M. However, we can still observe frequent but small constraint violations from 8~A.M. to 10~A.M., which then lead to overheating after 10 A.M. 

In comparison, the robust controller effectively handled these issues in an experiment conducted on $25^{th}$ May 2021. The result is shown in Figure~\ref{fig:exp_polydome_3}, where we used the same time varying indoor temperature constraint. We can observe that the robust controller safely protects the system from violating the lower bound through the whole test, and it also successfully pre-heated the building to fit the time-varying indoor temperature constraint. The performance deterioration that occurred to the non-robust controller after 10 A.M. was avoided as well, where the controller smoothly turned off the heating without unnecessary overheating.

 \definecolor{myblue}{rgb}{0.20, 0.6, 0.78}
\definecolor{mygreen}{rgb}{0.2,0.8,0.2}
\definecolor{myred}{rgb}{0.5,0,0}
 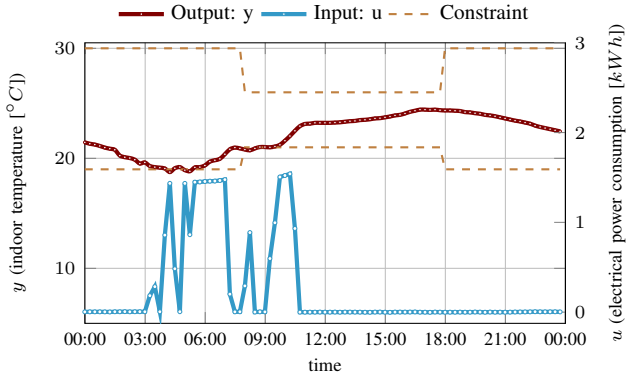
\begin{figure}
    \centering
    \begin{tikzpicture}
    \begin{axis}[
    date coordinates in=x,
    xmin=2021-05-13 23:59,
    xmax=2021-05-14 24:00,
    xtick distance=0.125,
    xticklabel=\hour:\minute,
    ymin=5, ymax= 30.5,
    enlargelimits=false,
    clip=true,
    grid=major,
    mark size=0.5pt,
    width=0.9\linewidth,
    height=0.6\linewidth,
    ylabel = {$y$ (indoor temperature [$^{\circ} C$])},
    xlabel= time,
    ylabel style={at={(axis description cs:0.05,0.5)}},
    xlabel style={at={(axis description cs:0.5,0.05)}},
    legend columns=3,
    label style={font=\scriptsize},
    ticklabel style = {font=\scriptsize},
    legend style={
    	font=\footnotesize,
    	draw=none,
		at={(0.5,1.03)},
        anchor=south
    }    
    ]
    
    \pgfplotstableread[col sep=comma]{data/polydome_min.dat}{\dat}

    \addplot+ [ultra thick,smooth, mark=*, mark options={fill=white, scale=1,line width = 0.1pt},myred] table [x={tt}, y={y}] {\dat};    
    \addlegendentry{Output: y}
    
    \addlegendimage{line legend,ultra thick,smooth, mark=*, mark options={fill=white, scale=1,line width = 0.2pt}, myblue}
    \addlegendentry{\change{Input}: u}       
    
    \addplot+ [thick,brown,dashed,mark = none] table [x={tt}, y={max}] {\dat};
    \addplot+ [thick,brown, dashed, mark = none] table [x={tt}, y={min}] {\dat};
    \addlegendentry{Constraint}     
    
    \end{axis}
    
    \begin{axis}[
    axis y line*=right,
    ymin=-0.125, ymax= 3,
    ylabel = {$u$ (electrical power consumption [$kWh$])},
    axis x line=none,
    date coordinates in=x,
    xmin=2021-05-13 23:59,
    xmax=2021-05-14 24:00,
    xtick distance=0.125,
    xticklabel=\hour:\minute,
    enlargelimits=false,
    mark size=0.5pt,
    width=0.9\linewidth,
    height=0.6\linewidth,
    legend style={
    	font=\footnotesize,
    	draw=none,
		at={(0.5,1.00)},
        anchor=south
    },
    ylabel style={at={(axis description cs:1.3,0.5)}},
    legend columns=2,
    label style={font=\scriptsize},
    ticklabel style = {font=\scriptsize}
    ]
    
    \pgfplotstableread[col sep=comma]{data/polydome_min.dat}{\dat}
   
    \addplot+ [ultra thick, mark=*, mark options={fill=white, scale=1.5,line width = 0.2pt}, myblue] table [x={tt}, y expr=\thisrow{u}*0.25] {\dat};
      
    \end{axis} 
    
    \end{tikzpicture}
    \caption{First experiment in \textit{Polydome}: one-day heating-mode running by the  bi-level data-driven  control without robust optimization }
    \label{fig:exp_polydome_2}     
 \end{figure}

 \definecolor{myblue}{rgb}{0.20, 0.6, 0.78}
\definecolor{mygreen}{rgb}{0.2,0.8,0.2}
\definecolor{myred}{rgb}{0.5,0,0}
 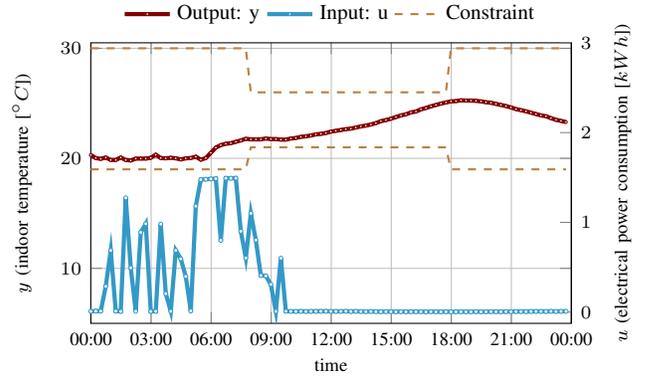
\begin{figure}
    \centering
    \begin{tikzpicture}
    \begin{axis}[
    date coordinates in=x,
    xmin=2021-05-24 23:59,
    xmax=2021-05-25 24:00,
    xtick distance=0.125,
    xticklabel=\hour:\minute,
    ymin=5, ymax= 30.5,
    enlargelimits=false,
    clip=true,
    grid=major,
    mark size=0.5pt,
    width=0.9\linewidth,
    height=0.6\linewidth,
    ylabel = {$y$ (indoor temperature [$^{\circ} C$])},
    xlabel= time,
    ylabel style={at={(axis description cs:0.05,0.5)}},
    xlabel style={at={(axis description cs:0.5,0.05)}},
    legend columns=3,
    label style={font=\scriptsize},
    ticklabel style = {font=\scriptsize},
    legend style={
    	font=\footnotesize,
    	draw=none,
		at={(0.5,1.03)},
        anchor=south
    }    
    ]
    
    \pgfplotstableread[col sep=comma]{data/polydome_min_robust.dat}{\dat}

    \addplot+ [ultra thick,mark=*, mark options={fill=white, scale=1,line width = 0.1pt},myred] table [x={tt}, y={y}] {\dat};    
    \addlegendentry{Output: y}
    
    \addlegendimage{line legend,ultra thick,smooth, mark=*, mark options={fill=white, scale=1,line width = 0.2pt}, myblue}
    \addlegendentry{\change{Input}: u}       
    
    \addplot+ [thick,brown,dashed,mark = none] table [x={tt}, y={max}] {\dat};
    \addplot+ [thick,brown, dashed, mark = none] table [x={tt}, y={min}] {\dat};
    \addlegendentry{Constraint}     
    
    \end{axis}
    
    \begin{axis}[
    axis y line*=right,
    ymin=-0.125, ymax= 3,
    ylabel = {$u$ (electrical power consumption [$kWh$])},
    axis x line=none,
    date coordinates in=x,
    xmin=2021-05-24 23:59,
    xmax=2021-05-25 24:00,
    xtick distance=0.125,
    xticklabel=\hour:\minute,
    enlargelimits=false,
    mark size=0.5pt,
    width=0.9\linewidth,
    height=0.6\linewidth,
    legend style={
    	font=\footnotesize,
    	draw=none,
		at={(0.5,1.00)},
        anchor=south
    },
    ylabel style={at={(axis description cs:1.3,0.5)}},
    legend columns=2,
    label style={font=\scriptsize},
    ticklabel style = {font=\scriptsize}
    ]
    
    \pgfplotstableread[col sep=comma]{data/polydome_min_robust.dat}{\dat}
   
    \addplot+ [ultra thick, mark=*, mark options={fill=white, scale=1.5,line width = 0.2pt}, myblue] table [x={tt}, y  expr=\thisrow{u}*0.25] {\dat};
      
    \end{axis} 
    
    \end{tikzpicture}
    \caption{First experiment in \textit{Polydome}: one-day heating-mode running by the  proposed  robust  data-driven  control}
    \label{fig:exp_polydome_3}        
 \end{figure}

\subsubsection{Experiment 2}
The second experiment is a pilot test to validate the adaptivity to the mode switching and the necessity of active excitation (Algorithm~\ref{alg:active_deepc}). The experiment was conducted from $28^{th}$ to $29^{th}$ May 2021, with the result shown in Figure~\ref{fig:exp_polydome_4}, where the indoor temperature of this experiment is bounded within
\begin{align*}
21\degree C\le y_i\le 25\degree C,&	
\end{align*}
This change of heating/cooling mode is depicted as a positive/negative control input and a red/blue shaded region in Figure~\ref{fig:exp_polydome_4}. However, it is noteworthy that the system was in cooling mode on the second evening. If there was no active excitation, the cooling should be off through the night to minimize energy consumption, and the Hankel matrices used for the controller would have lost persistent excitation. Instead, the active excitation, which is depicted as small fluctuations from 0:00 to 8:00 on the second day, maintained the persistency of excitation. The excitation signal is randomly selected as follows:
\begin{align*}
    \left\{
     \begin{array}{ll}
	 0 \,kWh \; \leq u_{e,i} \leq \;0.1\,kWh, &		\text{heating mode} \\
	 -0.075 \, kWh\;\leq u_{e,i} \leq \;0 \, kWh, &		\text{cooling mode}\\
	\end{array}
	\right.
\end{align*}
In conclusion, the controller successfully carried out the task of energy minimization in this experiment. In particular, when there is no need for heating/cooling, such as during the second evening, only active excitation took effect to maintain persistency of excitation. The heating/cooling also takes effect to robustly maintain the indoor temperature within the constraints.
 \definecolor{myblue}{rgb}{0.20, 0.6, 0.78}
\definecolor{mygreen}{rgb}{0.2,0.8,0.2}
\definecolor{myred}{rgb}{0.5,0,0}
 \definecolor{myblue2}{rgb}{0.94, 1, 1}
\definecolor{myred2}{rgb}{1,0.94,0.9}
 \begin{figure}
    \centering
    \begin{tikzpicture}
    \begin{axis}[
    date coordinates in=x,
    xmin=2021-05-27 20:59,
    xmax=2021-05-29 22:00,
    xtick distance=0.25,
    xticklabel=\hour:\minute,
    ymin=10, ymax= 26,
    enlargelimits=false,
    clip=true,
    grid=major,
    mark size=0.5pt,
    width=0.9\linewidth,
    height=0.62\linewidth,
    ylabel = {$y$ (indoor temperature [$^{\circ} C$])},
    xlabel= time,
    ylabel style={at={(axis description cs:0.05,0.5)}},
    xlabel style={at={(axis description cs:0.5,0.05)}},
    legend columns=4,
    label style={font=\scriptsize},
    ticklabel style = {font=\scriptsize},
    legend style={
    	font=\footnotesize,
    	draw=none,
		at={(0.5,1.03)},
        anchor=south
    }    
    ]
    
    \pgfplotstableread[col sep=comma]{data/polydome_min_robust_mode_color_v2.dat}{\dat}

    \addplot+ [ultra thick, mark=*, mark options={fill=white, scale=1,line width = 0.1pt},myred] table [x={tt}, y={y}] {\dat};    
    \addlegendentry{Output: y}
    
    \addlegendimage{line legend,ultra thick,smooth, mark=*, mark options={fill=white, scale=1,line width = 0.2pt}, myblue}
    \addlegendentry{\change{Input}: u}       
    
    \addplot+ [thick,brown,dashed,mark = none] table [x={tt}, y={max}] {\dat};
    \addlegendentry{Constraint}  
 
    \addlegendimage{line legend, thin, dashed, black, mark = none}
    \addlegendentry{\change{Active excitation}}     
    
    \addplot+ [thick,brown, dashed, mark = none] table [x={tt}, y={min}] {\dat};
      
    \end{axis}
    
    \begin{axis}[
    axis y line*=right,
    ymin=-1.5, ymax= 3,
    ylabel = {$u$ (electrical power consumption [$kWh$])},
    axis x line=none,
    date coordinates in=x,
    xmin=2021-05-27 20:59,
    xmax=2021-05-29 22:00,
    xtick distance=0.25,
    xticklabel=\hour:\minute,
    enlargelimits=false,
    mark size=0.5pt,
    width=0.9\linewidth,
    height=0.62\linewidth,
    legend style={
    	font=\footnotesize,
    	draw=none,
		at={(0.5,1.00)},
        anchor=south
    },
    ylabel style={at={(axis description cs:1.3,0.5)}},
    legend columns=2,
    label style={font=\scriptsize},
    ticklabel style = {font=\scriptsize}
    ]
    
    \pgfplotstableread[col sep=comma]{data/polydome_min_robust_mode_color_v2.dat}{\dat}
    
     \addplot+ [thin, dashed, black,mark = none] table [x={tt}, y  expr=\thisrow{if_du}*1] {\dat}; 
     
    \addplot+ [ultra thick, mark=*, mark options={fill=white, scale=1.5,line width = 0.2pt}, myblue] table [x={tt}, y  expr=\thisrow{u}*0.25] {\dat};

    \addplot+ [name path =A, ultra thin, myblue2,draw opacity=0.4,mark = none] table [x={tt}, y expr=\thisrow{mode_heat}*10] {\dat};
    \addplot+ [name path =B, ultra thin, myblue2,draw opacity=0.4,mark = none] table [x={tt}, y expr=\thisrow{mode_heat}*-10] {\dat};
    \addplot[myred,fill opacity =0.1] fill between[of=A and B]; 
    \addplot+ [name path =C, ultra thin, myred2,draw opacity=0.4,mark = none] table [x={tt}, y expr=\thisrow{mode_cool}*10] {\dat};
    \addplot+ [name path =D, ultra thin, myred2,draw opacity=0.4,mark = none] table [x={tt}, y expr=\thisrow{mode_cool}*-10] {\dat};
    \addplot[myblue,fill opacity=0.1] fill between[of=C and D]; 
    \end{axis} 
    
    \end{tikzpicture}
    \caption{Second experiment in \textit{Polydome}: two-day running by the  proposed  robust  data-driven  control. \change{The black line indicates the time interval within which the active extication is active: high level: active, low level: inactive.}}
    \label{fig:exp_polydome_4}        
 \end{figure}
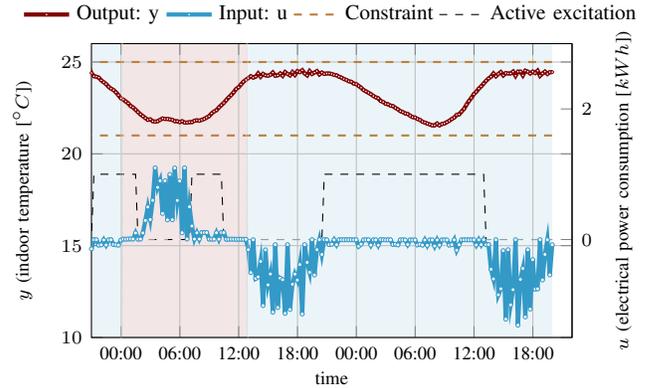

\subsubsection{Experiment 3}

This experiment was planned to validate the long-term reliability and adaptivity of the proposed controller. The experiment ran continuously for 20 days from $10^{th}$ June 2021 to $30^{th}$ June 2021. The statistics of this experiment is summarized in Table~\ref{tab:20d}, and we only plot results from the 20$^{th}$ 12:00 to the 30$^{th}$ 12:00 in Figure~\ref{fig:20d_2_output} to Figure~\ref{fig:20d_2_weather} due to the space limit~\footnote{Full plots of 20-day experiment are available on Github:\href{https://github.com/YingZhaoleo/Building_results}{https://github.com/YingZhaoleo/Building\_results}}. 

Note that the weather varies a lot throughout this experiment, with the outdoor temperature even once surpassing $29\degree C$ on the $20^{th}$ and also once dropping below $15\degree C$ on the $26^{th}$. Therefore, the experiment shows the adaptivity of the proposed controller to weather variation. Moreover, the cooling mode dominated the whole experiment, with only a few days of heating at night. The proposed controller gives a long-term guarantee of temperature constraint satisfaction (see Table~\ref{tab:20d}), while updating the Hankel matrices constantly. Regarding the data update, the active excitation scheme (Algorithm~\ref{alg:active_deepc}) also occasionally took effect to ensure persistency of excitation. \change{The time intervals within which the active excitation are active are plotted by value 1 along the black dashed line in Figure~\ref{fig:20d_2_output}.} 
\begin{table}[h!]
\begin{tabular}{ccccc}
 \shortstack{average\\ indoor\\ temperature}&  \shortstack{average \\ outdoor\\ temperature} & \shortstack{average\\solar\\radiation }&  \shortstack{duration of \\ constraint\\ violation } & \shortstack{average\\ hourly \change{energy}\\comsumption}  \\\hline
$23.7^\circ C$ &  $20.8^\circ C$ & $0.21W/m^2$ & $0h$ & $1.6kWh$\\\hline\hline
\shortstack{min/max\\indoor\\temperature}&\shortstack{min/max\\outdoor\\temperature}&\shortstack{min/max\\solar\\radiation}&\shortstack{hours of\\heating}&\shortstack{hours of \\cooling}\\\hline
$\substack{21.5^\circ C\\24.88^\circ C}$&$\substack{14.6^\circ C\\29.5^\circ C}$& $\substack{0W/m^2\\0.89W/m^2}$&$48h$& $432h$\\\hline
\end{tabular}
\caption{\label{tab:20d}Statistics of the 20-day experiment}
\end{table}

\subsubsection{Experiment 4}
Finally, we compare the proposed robust DeePC scheme with the default controller (Section.~\ref{sect:exp_setup}), which regulates the supply air temperature based on the return air temperature. A fixed setpoint at $22^\circ C$ is given to the default controller in order to ensure the occupants' comfort throughout the day. It is worth noting that the default controller is a benchmark controller widely used in the building control community~\cite{godina2018optimal,aswani2011reducing,yang2020model}.

The experiment with the proposed controller was conducted from $23^{rd}$ July 2021 to $26^{th}$ July 2021, and the one with the default controller ran from $7^{th}$ July 2021 to $10^{th}$ July 2021. To ensure a fair comparison, the weather conditions for these two experiments were similar, as shown in Figure~\ref{fig:polydome_compare_weather}. The indoor temperature and electrical power consumption are plotted in Figure~\ref{fig:polydome_compare_output}, alongside the statistics of these two experiments summarized in Table~\ref{tab:polydome_compare}.

From Table~\ref{tab:polydome_compare} and Figure~\ref{fig:polydome_compare_output}, more constraint violation is observed from the default controller than the proposed controller. One major underlying reason is that the default controller runs without the knowledge of a weather forecast and only cooled down the building when the return air temperature reached $23^\circ C$. We believe that is the reason accounting for the constraint violation in the day-1 experiment of the default controller. If the default controller could have predicted a high solar radiation and turned the cooling on constantly, the temperature constraint violation should have been avoided.

Besides the benefit of robust constraint satisfaction, the proposed controller is also more power efficient than the default controller under similar weather conditions, which in particular consumed $18.4 \%$ less electricity than the default controller (Table~\ref{tab:polydome_compare}). Note that, to maintain the data quality, the proposed controller ran the active excitation scheme (Algorithm~\ref{alg:active_deepc}) regularly from 00:00 to 9:00. Thus, there is still the possibility to further improve the energy efficiency of the proposed controller.

\begin{table}[h!]
\begin{tabular}{ccccc}
 & \shortstack{average\\ indoor\\ temperature}&  \shortstack{average \\ outdoor\\ temperature} & \shortstack{average\\solar\\radiation }&  \shortstack{duration of \\ constraint\\ violation }  \\\hline
\shortstack{Proposed \\controller} & $23.1^\circ C$ &  $18.5^\circ C$ & $0.17W/m^2$ & $0h$ \\\hline
\shortstack{Default \\controller} & $23.2^\circ C$ &  $18.8^\circ C$ & $0.17W/m^2$ & $5.75h$ \\\hline\hline
& \shortstack{min/max\\indoor\\temperature}&\shortstack{min/max\\outdoor\\temperature}&\shortstack{min/max\\solar\\radiation}& \shortstack{averaged\\ hourly \change{energy}\\comsumption} \\\hline
\shortstack{Proposed \\controller}  & $\substack{21.5^\circ C\\24.9^\circ C}$&$\substack{15.2^\circ C\\23.4^\circ C}$& $\substack{0W/m^2\\0.80W/m^2}$&$1.15kWh$\\ \hline
\shortstack{Default \\controller}  & $\substack{21.2^\circ C\\25.9^\circ C}$&$\substack{15.5^\circ C\\24.1^\circ C}$& $\substack{0W/m^2\\0.92W/m^2}$&$1.41kWh$\\ \hline
\end{tabular}
\caption{\label{tab:polydome_compare}Statistics of the four-day comparison}
\end{table}

\section{Conclusion}\label{sect:conclusion}
In this work, we propose a robust bi-level data-driven adaptive-predictive building controller based on the Willems' fundamental lemma. The proposed scheme performs comparably to a classical model-based approach in a numerical simulation. The practical viability is shown by deploying the proposed scheme to a conference building on the EPFL campus, where, without extra modelling effort, our proposed scheme improve 18\% energy efficiency and robustly ensures occupant comfort.

{
\bibliographystyle{abbrv}
\bibliography{ref.bib}}

\clearpage

 \definecolor{myblue}{rgb}{0.20, 0.6, 0.78}
\definecolor{mygreen}{rgb}{0.2,0.8,0.2}
\definecolor{myred}{rgb}{0.5,0,0}
 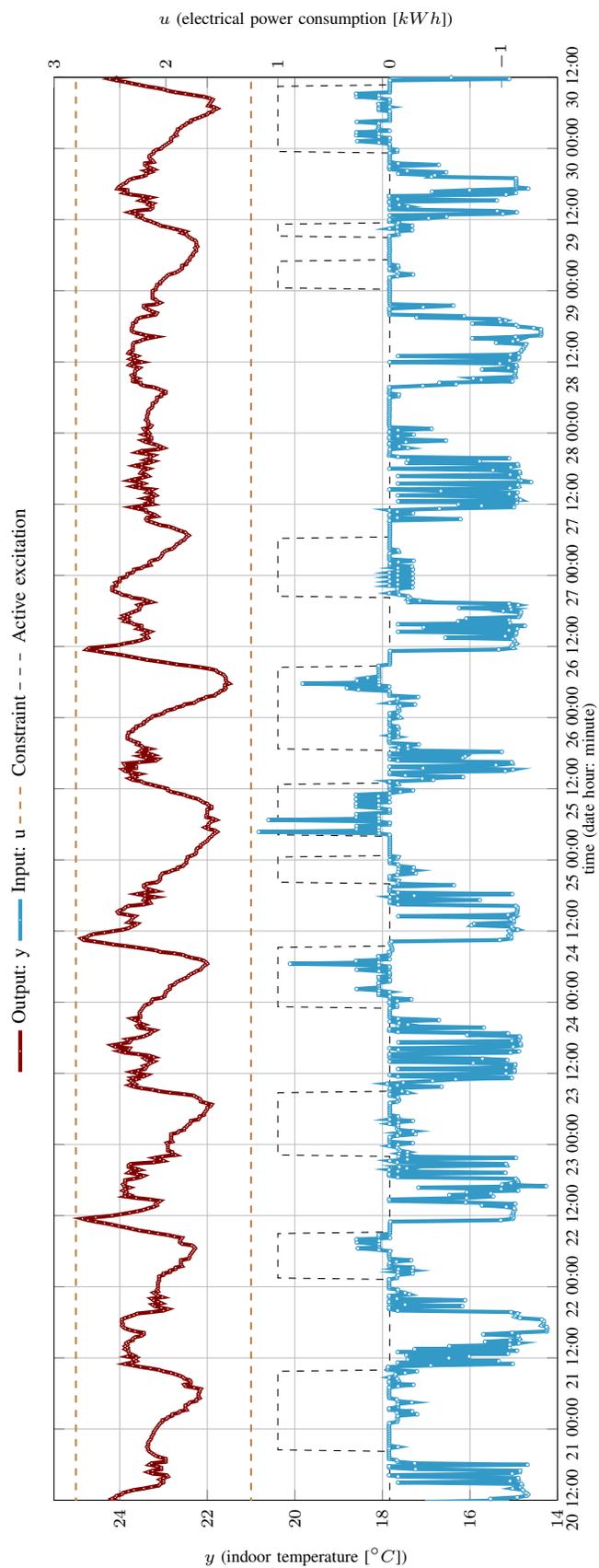
\begin{figure}
    \centering
    \begin{tikzpicture}[rotate=90,transform shape]
    \begin{axis}[
    date coordinates in=x,
    xmin=2021-06-20 12:00,
    xmax=2021-06-30 12:00,
    xtick distance=0.5,
    xticklabel=\day\ \hour:\minute,
    ymin= 14, ymax= 25.5,
    enlargelimits=false,
    clip=true,
    grid=major,
    mark size=0.5pt,
    width=2.5\linewidth,
    height=\linewidth,
    ylabel = {$y$ (indoor temperature [$^{\circ} C$])},
    xlabel= {time (date hour: minute)},
    ylabel style={at={(axis description cs:0.02,0.5)},rotate=180},
    xlabel style={at={(axis description cs:0.5,0.04)}},
    legend columns=4,
    label style={font=\scriptsize},
    ticklabel style = {font=\scriptsize},
    legend style={
    	font=\footnotesize,
    	draw=none,
		at={(0.5,1.03)},
        anchor=south
    }    
    ]
    
    \pgfplotstableread[col sep=comma]{data/20day_2.dat}{\dat}

    \addplot+ [ultra thick, mark=*, mark options={fill=white, scale=1,line width = 0.1pt},myred] table [x={t}, y={y}] {\dat};    
    \addlegendentry{Output: y}
    
    \addlegendimage{line legend,ultra thick,smooth, mark=*, mark options={fill=white, scale=1,line width = 0.2pt}, myblue}
    \addlegendentry{Input: u} 
    
    \addplot+ [thick,brown,dashed,mark = none,forget plot] table [x={t}, y={max}] {\dat};
    \addplot+ [thick,brown, dashed, mark = none] table [x={t}, y={min}] {\dat};
    \addlegendentry{Constraint}

    \addlegendimage{thin, dashed,mark = none, black}
    \addlegendentry{\change{Active excitation}}

    \end{axis}
    
    \begin{axis}[
    axis y line*=right,
    ymin=-1.5, ymax= 3,
    ylabel = {$u$ (electrical power consumption [$kWh$])},
    axis x line=none,
    date coordinates in=x,
    xmin=2021-06-20 12:00,
    xmax=2021-06-30 12:00,
    xtick distance=0.5,
    xticklabel=\day\ \hour:\minute,
    enlargelimits=false,
    clip=true,
    grid=none,
    mark size=0.5pt,
    width=2.5\linewidth,
    height=\linewidth,
    ylabel style={at={(axis description cs:1.1,0.5)},rotate=180},
    legend columns=3,
    label style={font=\scriptsize},
    ticklabel style = {font=\scriptsize},
    legend style={
    	font=\footnotesize,
    	draw=none,
		at={(0.5,1.03)},
        anchor=south
    }    
    ]
    
    \pgfplotstableread[col sep=comma]{data/20day_2_v2.dat}{\dat}
    
    \addplot+ [thin, dashed,mark = none, black] table [x={t}, y  expr=\thisrow{if_du}*1] {\dat}; 

    \addplot+ [ultra thick, mark=*, mark options={fill=white, scale=1.5,line width = 0.2pt}, myblue] table [x={t}, y expr=\thisrow{u}*0.25] {\dat};

    \end{axis}
    
    \end{tikzpicture}
    \caption{Experimental results from noon $20^{th}$ June 2021 to noon $30^{th}$ June 2021: indoor temperature and electrical power consumption. \change{The black line indicates the time interval within which the active extication is active: high level: active, low level: inactive.} \label{fig:20d_2_output}}
\end{figure}

 \begin{figure}
    \centering
    \begin{tikzpicture}[rotate=90,transform shape]
    \begin{axis}[
    date coordinates in=x,
    xmin=2021-06-20 12:00,
    xmax=2021-06-30 12:00,
    xtick distance=0.5,
    xticklabel=\day\ \hour:\minute,
    ymin= 13, ymax= 28,
    enlargelimits=false,
    clip=true,
    grid=major,
    mark size=0.5pt,
    width=2.5\linewidth,
    height=\linewidth,
    ylabel = {$w_1$ (outdoor temperature [$^{\circ} C$])},
    xlabel= {time (date hour: minute)},
    ylabel style={at={(axis description cs:0.02,0.5)},rotate=180},
    xlabel style={at={(axis description cs:0.5,0.04)}},
    legend columns=3,
    label style={font=\scriptsize},
    ticklabel style = {font=\scriptsize},
    legend style={
    	font=\footnotesize,
    	draw=none,
		at={(0.5,1.03)},
        anchor=south
    }    
    ]
    
    \pgfplotstableread[col sep=comma]{data/20day_2.dat}{\dat}

    \addplot+ [ultra thick, mark=*, mark options={fill=white, scale=1,line width = 0.1pt},opacity=0.8,myred] table [x={t}, y={w1}] {\dat};    
    \addlegendentry{Outdoor temperature} 
    \addlegendimage{line legend,ultra thick,smooth, mark=*, mark options={fill=white, scale=1,line width = 0.2pt}, mygreen}
    \addlegendentry{Solar radiation}

    \end{axis}
     \begin{axis}[
    axis y line*=right,
    ymin=-0.5, ymax= 6,
    ylabel = {$w_2$ (solar radiation [$kW/m^2$])},
    axis x line=none,
    date coordinates in=x,
    xmin=2021-06-20 12:00,
    xmax=2021-06-30 12:00,
    xtick distance=0.5,
    xticklabel=\day\ \hour:\minute,
    enlargelimits=false,
    clip=true,
    grid=none,
    mark size=0.5pt,
    width=2.5\linewidth,
    height=\linewidth,
    ylabel style={at={(axis description cs:1.1,0.5)},rotate=180},
    legend columns=3,
    label style={font=\scriptsize},
    ticklabel style = {font=\scriptsize},
    legend style={
    	font=\footnotesize,
    	draw=none,
		at={(0.5,1.03)},
        anchor=south
    }    
    ]
    
    \pgfplotstableread[col sep=comma]{data/20day_2.dat}{\dat}

    \addplot+ [ultra thick, mark=*, mark options={fill=white, scale=1.5,line width = 0.2pt}, mygreen] table [x={t}, y={w2}] {\dat};

    \end{axis}
    \end{tikzpicture}
    \caption{Experimental results from noon $20^{th}$ June 2021 to noon $30^{th}$ June 2021: outdoor temperature and solar radiation\label{fig:20d_2_weather}\\ \\}
\end{figure}
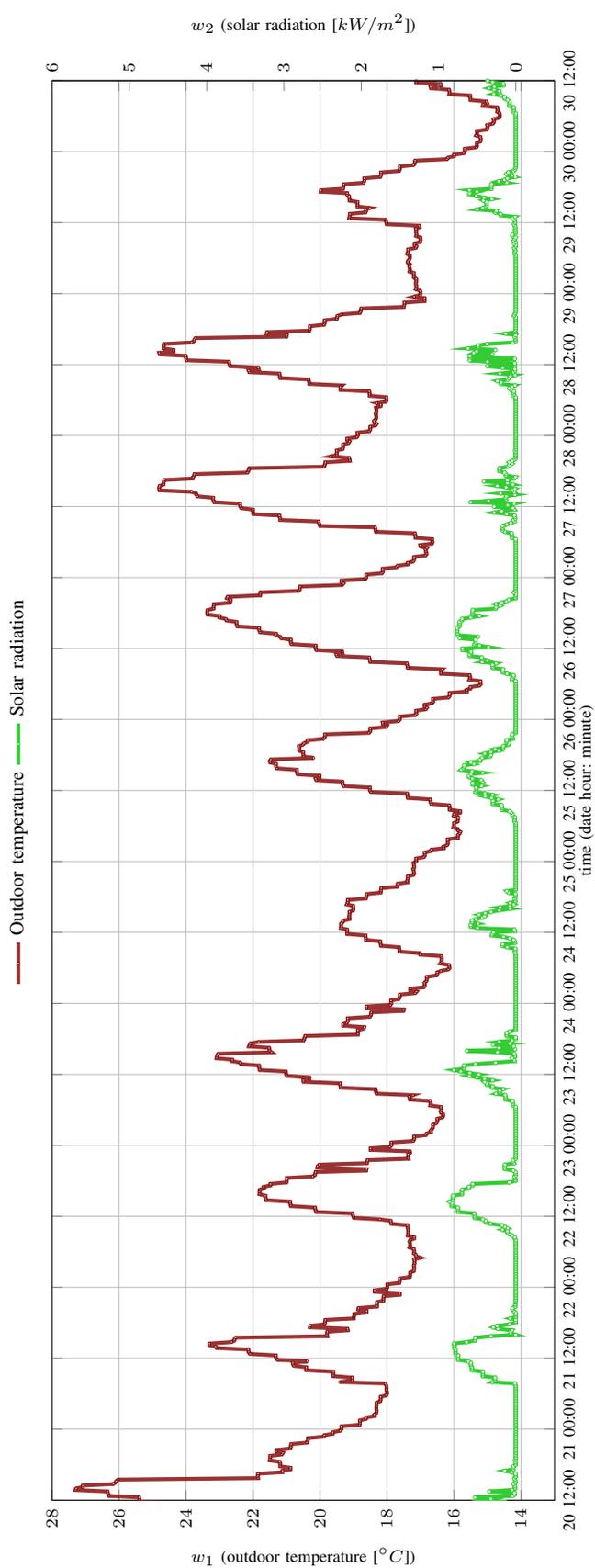
 \definecolor{myblue}{rgb}{0.20, 0.6, 0.78}
\definecolor{mygreen}{rgb}{0.2,0.8,0.2}
\definecolor{myred}{rgb}{0.5,0,0}
\definecolor{myorange}{rgb}{1,0.6,0.07}
 \begin{figure}
    \centering
    \begin{tikzpicture}[rotate=90,transform shape]
    \begin{axis}[
    date coordinates in=x,
    xmin=2021-07-06 23:45,
    xmax=2021-07-11 03:30,
    xtick distance=0.25,
    xticklabel=\hour:\minute,
    ymin= 14, ymax= 26.5,
    enlargelimits=false,
    clip=true,
    grid=major,
    mark size=0.5pt,
    width=2.5\linewidth,
    height=\linewidth,
    ylabel = {$y$ (indoor temperature [$^{\circ} C$])},
    xlabel= {time (date hour: minute)},
    ylabel style={at={(axis description cs:0.02,0.5)},rotate=180},
    xlabel style={at={(axis description cs:0.5,0.04)}},
    legend columns=3,
    label style={font=\scriptsize},
    ticklabel style = {font=\scriptsize},
    legend style={
    	font=\footnotesize,
    	draw=none,
		at={(0.5,1.03)},
        anchor=south
    }    
    ]
    
    \pgfplotstableread[col sep=comma]{data/polydome_compare_4day.dat}{\dat}

    \addplot+ [ultra thick, mark=*, mark options={fill=white, scale=1,line width = 0.1pt},myred] table [x={tt_rule}, y={y_deepc}] {\dat};    
    \addlegendentry{Proposed controller: output y}
    
    \addplot+ [ultra thick, mark=*, mark options={fill=white, scale=1,line width = 0.1pt},mygreen] table [x={tt_rule}, y={y_rule}] {\dat};    
    \addlegendentry{Default controller: output y}  
    
    \addplot+ [thick,brown,dashed,mark = none] table [x={tt_rule}, y={max}] {\dat};
    \addlegendentry{Constraint}    
    
    \addlegendimage{line legend,ultra thick,smooth, mark=*, mark options={fill=white, scale=1,line width = 0.2pt}, myblue}
    \addlegendentry{Proposed controller: input u} 
    \addlegendimage{line legend,ultra thick,smooth, mark=*, mark options={fill=white, scale=1,line width = 0.2pt}, myorange}
    \addlegendentry{Default controller: input u} 
    
    \addplot+ [thick,brown, dashed, mark = none] table [x={tt_rule}, y={min}] {\dat};

    \end{axis}
    
    \begin{axis}[
    axis y line*=right,
    ymin=-2.0, ymax= 4.5,
    ylabel = {$u$ (electrical power consumption [$kWh$])},
    axis x line=none,
    date coordinates in=x,
    xmin=2021-07-06 23:45,
    xmax=2021-07-11 03:30,
    xtick distance=0.25,
    xticklabel= \hour:\minute,
    enlargelimits=false,
    clip=true,
    grid=none,
    mark size=0.5pt,
    width=2.5\linewidth,
    height=\linewidth,
    ylabel style={at={(axis description cs:1.1,0.5)},rotate=180},
    legend columns=3,
    label style={font=\scriptsize},
    ticklabel style = {font=\scriptsize},
    legend style={
    	font=\footnotesize,
    	draw=none,
		at={(0.5,1.03)},
        anchor=south
    }    
    ]
    
    \pgfplotstableread[col sep=comma]{data/polydome_compare_4day.dat}{\dat}

    \addplot+ [ultra thick, mark=*, mark options={fill=white, scale=1.5,line width = 0.2pt}, myblue] table [x={tt_rule}, y expr=\thisrow{u_deepc}*-0.25] {\dat};  
    \addplot+ [ultra thick, mark=*, mark options={fill=white, scale=1.5,line width = 0.2pt}, myorange] table [x={tt_rule}, y expr=\thisrow{u_rule}*-0.25] {\dat};  

    \end{axis}
    
    \end{tikzpicture}
    \caption{Comparison  of  the  proposed  robust  controller  and  default controller: indoor temperature and electrical power consumption \label{fig:polydome_compare_output}}
\end{figure}
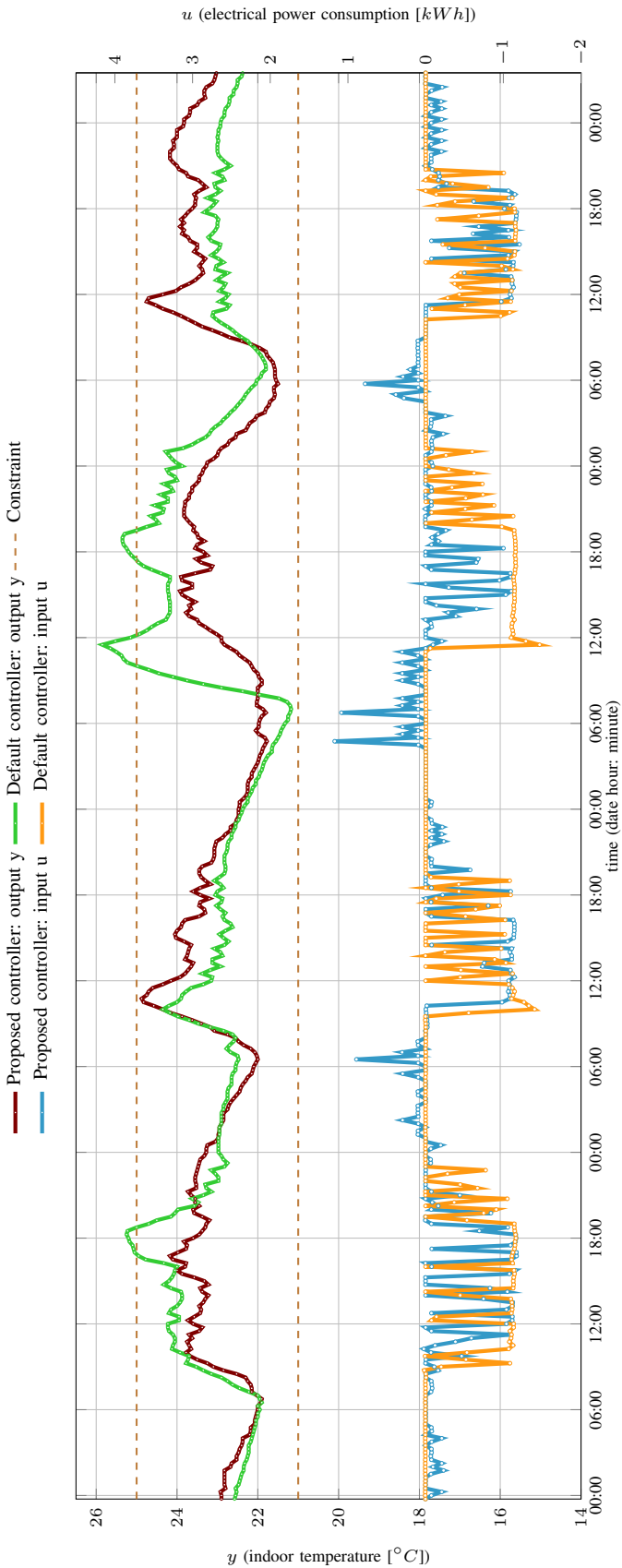

 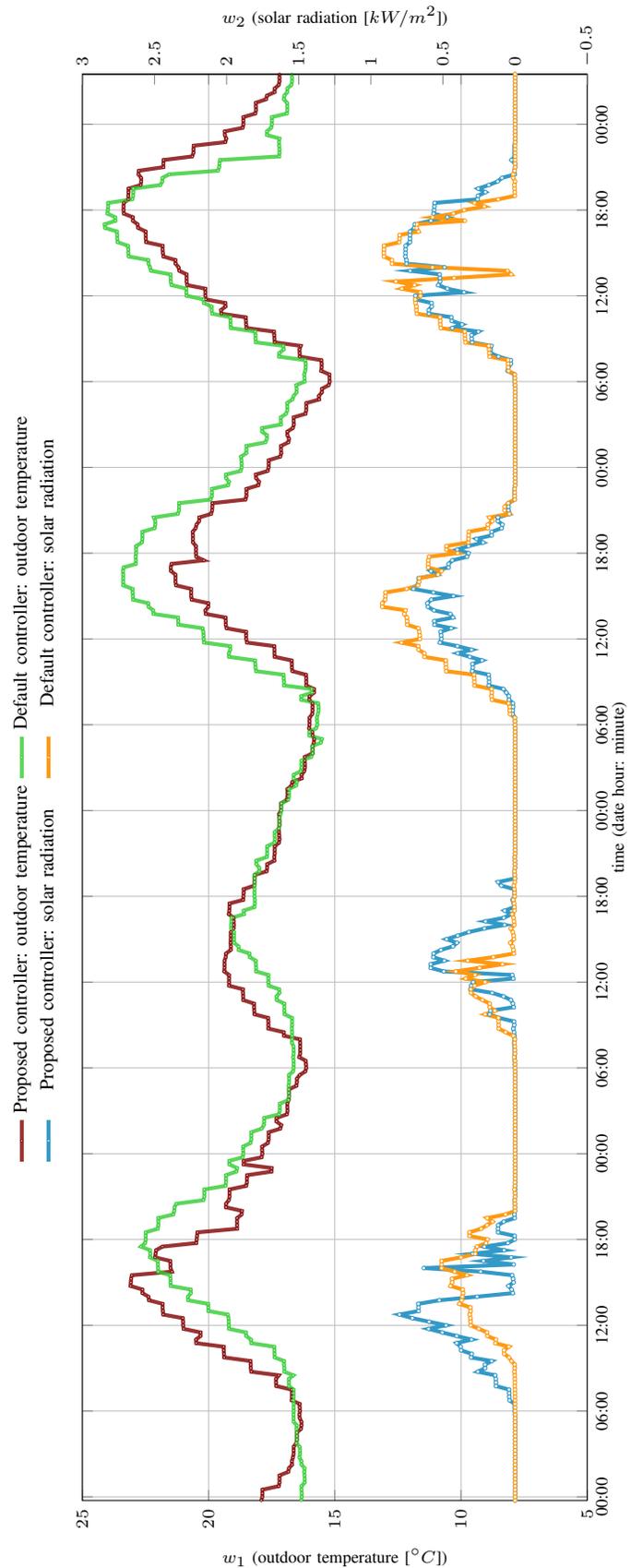
\begin{figure}
    \centering
    \begin{tikzpicture}[rotate=90,transform shape]
    \begin{axis}[
    date coordinates in=x,
    xmin=2021-07-06 23:45,
    xmax=2021-07-11 03:30,
    xtick distance=0.25,
    xticklabel= \hour:\minute,
    ymin= 5, ymax= 25,
    enlargelimits=false,
    clip=true,
    grid=major,
    mark size=0.5pt,
    width=2.5\linewidth,
    height=\linewidth,
    ylabel = {$w_1$ (outdoor temperature [$^{\circ} C$])},
    xlabel= {time (date hour: minute)},
    ylabel style={at={(axis description cs:0.02,0.5)},rotate=180},
    xlabel style={at={(axis description cs:0.5,0.04)}},
    legend columns=2,
    label style={font=\scriptsize},
    ticklabel style = {font=\scriptsize},
    legend style={
    	font=\footnotesize,
    	draw=none,
		at={(0.5,1.03)},
        anchor=south
    }    
    ]
    
    \pgfplotstableread[col sep=comma]{data/polydome_compare_4day.dat}{\dat}

    \addplot+ [ultra thick, mark=*, mark options={fill=white, scale=1,line width = 0.1pt},opacity=0.8,myred] table [x={tt_rule}, y={w1_deepc}] {\dat};    
    \addlegendentry{Proposed  controller: outdoor temperature} 
    \addplot+ [ultra thick, mark=*, mark options={fill=white, scale=1,line width = 0.1pt},opacity=0.8,mygreen] table [x={tt_rule}, y={w1_rule}] {\dat};    
    \addlegendentry{Default controller: outdoor temperature} 
    
    \addlegendimage{line legend,ultra thick,smooth, mark=*, mark options={fill=white, scale=1,line width = 0.2pt}, myblue}
    \addlegendentry{Proposed  controller: solar radiation} 
    \addlegendimage{line legend,ultra thick,smooth, mark=*, mark options={fill=white, scale=1,line width = 0.2pt}, myorange}
    \addlegendentry{Default controller: solar radiation}

    \end{axis}
     \begin{axis}[
    axis y line*=right,
    ymin=-0.5, ymax= 3,
    ylabel = {$w_2$ (solar radiation [$kW/m^2$])},
    axis x line=none,
    date coordinates in=x,
    xmin=2021-07-06 23:45,
    xmax=2021-07-11 03:30,
    xtick distance=0.25,
    xticklabel= \hour:\minute,
    enlargelimits=false,
    clip=true,
    grid=none,
    mark size=0.5pt,
    width=2.5\linewidth,
    height=\linewidth,
    ylabel style={at={(axis description cs:1.1,0.5)},rotate=180},
    legend columns=3,
    label style={font=\scriptsize},
    ticklabel style = {font=\scriptsize},
    legend style={
    	font=\footnotesize,
    	draw=none,
		at={(0.5,1.03)},
        anchor=south
    }    
    ]
    
    \pgfplotstableread[col sep=comma]{data/polydome_compare_4day.dat}{\dat}

    \addplot+ [ultra thick, mark=*, mark options={fill=white, scale=1.5,line width = 0.2pt}, myblue] table [x={tt_rule}, y={w2_deepc}] {\dat};  
    \addplot+ [ultra thick, mark=*, mark options={fill=white, scale=1.5,line width = 0.2pt}, myorange] table [x={tt_rule}, y={w2_rule}] {\dat};

    \end{axis}
    \end{tikzpicture}
    \caption{Comparison  of  the  proposed  robust  controller  and  default controller: outdoor temperature and solar radiation\label{fig:polydome_compare_weather}}
\end{figure}

\end{document}